\documentclass[fleqn,usenatbib]{mnras}

\usepackage[T1]{fontenc}
\usepackage{ae,aecompl}
\usepackage{times}
\usepackage{graphicx}	% Including figure files
\usepackage{amsmath}	% Advanced maths commands
\usepackage{amssymb}	% Extra maths symbols
\usepackage{soul}
\usepackage[normalem]{ulem}
\usepackage{xcolor}
\usepackage{xspace}
\usepackage{scalerel}
\usepackage{tikz}
\usetikzlibrary{svg.path}

\definecolor{orcidlogocol}{HTML}{A6CE39}
\tikzset{
  orcidlogo/.pic={
    \fill[orcidlogocol] svg{M256,128c0,70.7-57.3,128-128,128C57.3,256,0,198.7,0,128C0,57.3,57.3,0,128,0C198.7,0,256,57.3,256,128z};
    \fill[white] svg{M86.3,186.2H70.9V79.1h15.4v48.4V186.2z}
                 svg{M108.9,79.1h41.6c39.6,0,57,28.3,57,53.6c0,27.5-21.5,53.6-56.8,53.6h-41.8V79.1z M124.3,172.4h24.5c34.9,0,42.9-26.5,42.9-39.7c0-21.5-13.7-39.7-43.7-39.7h-23.7V172.4z}
                 svg{M88.7,56.8c0,5.5-4.5,10.1-10.1,10.1c-5.6,0-10.1-4.6-10.1-10.1c0-5.6,4.5-10.1,10.1-10.1C84.2,46.7,88.7,51.3,88.7,56.8z};
  }
}

\newcommand\orcidicon[1]{\href{https://orcid.org/#1}{\mbox{\scalerel*{
\begin{tikzpicture}[yscale=-1,transform shape]
\pic{orcidlogo};
\end{tikzpicture}
}{|}}}}

\newcommand{\rr}{\mathbf{r}}
\newcommand{\dirac}{\delta_{\rm D}}
\newcommand{\heaviside}{\Theta_{\rm H}}

\newcommand{\vvec}[1]{\mathbf{#1}}
 \newcommand{\dd}{\mathrm{d}}

\newcommand{\ie}{\textsl{i.e.}\xspace}
\newcommand{\eg}{e.g.\xspace}
\newcommand{\mpcph}{\,$h^{-1}$Mpc}

\newcommand{\add}[1]{\protect\hl{#1} }

 \title[Clustering of critical points]{The clustering of critical points in the evolving cosmic web}
\author[J.~Shim, S.~Codis, C.~Pichon, D.~Pogosyan and C.~Cadiou]{
J.~Shim$^{\protect\orcidicon{0000-0001-7352-6175}1}$\thanks{E-mail: jsshim@kias.re.kr},
S.~Codis$^{2,3}$,
C.~Pichon$^{\protect\orcidicon{0000-0003-0695-6735}1,2,3}$,
D.~Pogosyan$^{1,4}$
and
C.~Cadiou$^{\protect\orcidicon{0000-0003-2285-0332}5}$
\\
$^{1}$ Korea Institute for Advanced Study, 85 Hoegiro, Dongdaemun-gu, 02455 Seoul, Republic of Korea\\
$^{2}$ CNRS and Sorbonne Universit\'e, UMR 7095, Institut d'Astrophysique de Paris, 98 bis Boulevard Arago, 75014 Paris, France\\
$^{3}$ IPhT, DRF-INP, UMR 3680, CEA, Orme des Merisiers Bat 774, 91191 Gif-sur-Yvette, France\\
$^{4}$ Department of Physics, University of Alberta, 11322-89 Avenue, Edmonton, Alberta, T6G 2G7, Canada\\
$^{5}$ Department of Physics and Astronomy, University College London, London WC1E 6BT, United Kingdom\\
}

\date{\today}
\pubyear{2020}

\usepackage{hyperref} %<--- Load after everything else

\begin{document}
\label{firstpage}
\pagerange{\pageref{firstpage}--\pageref{lastpage}}
\maketitle

% Abstract of the paper
\begin{abstract}
Focusing on both small separations and Baryonic Acoustic Oscillation scales, the cosmic evolution of the  clustering properties of peak, void, wall and filament-type  critical points  is measured using two-point correlation functions  in $\Lambda$CDM dark matter simulations as a function of their relative rarity. A qualitative comparison to the corresponding theory for Gaussian Random fields allows us to understand the following observed features: i) the appearance of an exclusion zone  at small separation, whose size depends both on rarity and  on the signature (\ie the number of negative eigenvalues) of the  critical points involved;  ii) the amplification of the Baryonic Acoustic Oscillation bump with rarity and its reversal for cross correlations involving negatively biased critical points; iii)  the  orientation-dependent small-separation divergence of the cross-correlations of peaks and filaments (resp. voids and walls) which reflects the relative  loci of such points in the filament's (resp. wall's) eigenframe. The most significant features of the correlations are tabulated.
The (cross-) correlations involving the most non-linear critical points (peaks, voids) display significant variation with redshift, while those involving less non-linear critical points seem mostly insensitive to redshift evolution, which should prove advantageous to model. The relative distances to the maxima of the peak-to-wall and peak-to-void over that of the peak-to-filament cross-correlation are in ratios of $\sim\sqrt{2}$ and $\sim\sqrt{3}$, respectively which could be interpreted as an indication of the cosmic crystal being on average close to a cubic lattice. The insensitivity to redshift evolution suggests that the absolute and relative clustering of critical points could become a topologically robust alternative to standard clustering techniques when analysing upcoming large scale surveys such as Euclid or LSST.
\end{abstract}


%%%%%%%%%%%%%%%%% BODY OF PAPER %%%%%%%%%%%%%%%%%%

%%%%%%%%%%%%%%%%%%%%%%%%%%%%%%%%%%%%%%%%%%%%%%%%%%
\section{Introduction}\label{sec:intro}
%%%%%%%%%%%%%%%%%%%%%%%%%%%%%%%%%%%%%%%%%%%%%%%%%%

The clustering properties of  the cluster and galaxy distribution in extra-galactic surveys has been historically a major source of information for cosmology  \citep{1970AJ.....75...13P,1974ApJ...187..425P,1981ApJ...243L.119P,1983ApJ...270...20B,1986MNRAS.222..323K,1991ApJ...383..104K, 2005RvMP...77..207V,2009ApJ...692..265E,2016MNRAS.458.1909V,2018A&A...620A...1M}. The $N$-point functions of the galaxies themselves has received most attention
\citep{2003MNRAS.346...78H,2005MNRAS.364..620G,2007MNRAS.378.1196K,2015MNRAS.452.1914G}. 
But more recently, with larger upcoming spectroscopic or photometric surveys\footnote{
\href{https://www.euclid-ec.org}{Euclid},
\href{http://www.lsst.org}{LSST},
\href{https://www.nasa.gov/content/goddard/nancy-grace-roman-space-telescope}{WFirst},
\href{https://spherex.caltech.edu}{SphereX},
%\href{https://www.the-athena-x-ray-observatory.eu}{Athena}, 
\href{https://ingconfluence.ing.iac.es:8444/confluence//display/WEAV/The+WEAVE+Project}{WEAVE},
\href{https://www.cfht.hawaii.edu/en/news/MSE-new/}{MSE},
\href{https://pfs.ipmu.jp}{PFS},
to name a few.
}, the specific clustering of special features like peaks   \citep{PeacockHeavens,Regos1995,matsubara&codis19} and voids have drawn some interest  \citep{2014PhRvD..90j3521C, 2016MNRAS.459.4020L, 2017arXiv171201002L,2017MNRAS.468.4822L}.
Voids in particular are underdense regions, which make them sensitive to physics beyond the standard model while evolving in a quasi-linear regime \citep{hamaus_ProbingCosmologyGravity_2015,stopyra_HowBuildCatalogue_2020}. They offer a unique environment to test alternative theories of gravity or dark energy 
\citep{2013PhRvL.111x1103S,2015JCAP...08..028B,2017PhRvD..95b4018V,2018PhRvD..98b3511B}.
Massive neutrinos are also known to impact the properties of voids, be it their bias \citep{schuster_BiasCosmicVoids_2019}, their size distribution \citep{2015JCAP...11..018M,2016JCAP...11..015B}, or their two-point correlation functions \citep{2019MNRAS.488.4413K}.

In this paper our  main  objective is to extend those investigation towards the cosmic evolution of the 
full set of components of the Cosmic Web \citep{bkp96} focusing on auto-correlation functions of filaments and walls on the one hand, and cross-correlation of voids, peaks, walls and filaments on the other.
Indeed, this  subset of special points within the density field have remarkable topological properties which makes them more robust to  systematics, while their Lagrangian two-point statistics can be computed from first principle, making explicit their  (initial) dependency on the underlying cosmological parameters.

We will in particular focus on the  Baryonic Acoustic Oscillation (BAO) scale,  since it is one of the  most prominent signatures of  two-point functions \citep{2007PhRvD..76f3009M,2008PhRvD..77b3533C,desjacques08} which has now been clearly identified in surveys \citep{2010MNRAS.401.2148P,2011MNRAS.418.1707B,2011MNRAS.416.3017B,2014JCAP...05..027F} and used to infer cosmological parameters.  We  aim to  study the redshift evolution of the exclusion zone at smaller scale \citep{baldauf+16,Desjacques2018}, together with the  shape of the two-point correlation functions (2pCFs)  near their maximum. This should prove to be a useful dark energy probe, since \eg  2pCFs of walls  inform us about the characteristic size of voids whose redshift evolution provides information on cosmological parameters \citep{li+12,cai+15,hamaus_ProbingCosmologyGravity_2015,pisani+15,2016PhRvL.117i1302H,hamaus_MultipoleAnalysisRedshiftspace_2017,verza+19}.
We will also study the angular dependence of the cross-correlations involving saddle points as measured in their eigen-frame, that is to say the oriented stacking of extrema (voids or peaks) in the frame of a saddle-point understood as the middle of a filament. 
We will also investigate the redshift evolution of those cross-correlation function together with their dependence with the rarity of the critical points.

Section~\ref{sec:method} presents the simulations and the estimators used to identify critical points and 
compute their correlations.
Section~\ref{sec:measurements} presents number density counts and  clustering measurements in the simulation's snapshots,
Section~\ref{sec:crystalMain} identifies the pseudo cubic lattice structure of the cosmic web,
while Section~\ref{sec:conclusions} concludes.

Appendix~\ref{sec:theory} presents theoretical predictions for Gaussian random fields, while Appendix~\ref{sec:cross-rare} discusses briefly the divergence at small separation of the cross-correlations between peaks and filaments (resp. wall and voids).

For clarity, we have systematically presented  in the main text
all auto- and cross-correlations measured  in the simulation as a function of  abundance and 
redshift, and postponed the theoretical correlations  for Gaussian random fields to the Appendices,
to which we refer on a case-by-case basis in the main text.
Tables~\ref{tab:peak_posh_rarity}--\ref{tab:exclu_bao_redshift} present a summary of the typical scales and amplitudes
characterising these correlations.
%%%%%%%%%%%%%%%%%%%%%%%%%%%%%%%%%%%%%%%%%%%%%%%%%%
\section{Method}\label{sec:method}
%%%%%%%%%%%%%%%%%%%%%%%%%%%%%%%%%%%%%%%%%%%%%%%%%%
Let us first describe the simulations we used to compute the position of critical points before presenting the corresponding clustering estimator. 

\subsection{Simulations}
 We have run 532 vanilla $\Lambda$CDM simulations in a 500\mpcph\ box using the cosmological parameters  $\Omega_{\rm b}=0.04$, $\Omega_{\rm m}=0.24$, $\Omega_{\Lambda}=0.76$, $h=0.70$, $\sigma_8=0.8$, $n_{s}=1$.  Each simulation follows  $256^3$ particles down to redshift zero using {\sc Gadget2} \citep{Gadget2001}. At each snapshot, the density field is sampled on a $256^3$ grid smoothed with a Gaussian filter over  $R_{\rm G}={6}$ \mpcph.
We identified the range of redshifts not impacted by transients in the simulation ($z\leq 4$)
by computing ratios of the expectation of higher order cumulants of the field (commonly referred to as reduced cumulants) and comparing to expectation from tree-order 
perturbation theory. We binned the snapshots in the range of $0\le z<4$ into five sub-samples, three
with $\Delta z=1$ for $ 4 < z \le 1$, $0.5 \le z < 1$ and $0 \le z < 0.5$. The corresponding root mean square (rms) fluctuation of the density field averaged over the snapshots in each redshift bin increases with time from $\sigma=0.18$ ($3\le z<4$) to $\sigma=0.56$ ($0\le z<0.5$) and allows us to probe the mildly non-linear regime of structure formation which is the sweet spot for extracting cosmological information thanks to higher order statistics beyond the weakly non-linear regime which can be captured from first principle calculations.
Note that this set of simulations was also used in \cite{cadiou2020} to study
critical events corresponding to the mergers of critical points.

\subsection{Estimators for critical statistics}

\subsubsection{Critical points definition}
The critical points of a field are points where the gradient vanishes \citep{milnor1963morse,BBKS,arnold2006}. There exist different types of critical points that are classified by the sign of the eigenvalues of the hessian matrix (the matrix of the second derivatives of the field) at that point. Based on the typical shape of the isosurfaces in their neighbourhood, we typically call the four types of critical points in the 3D cosmic density field as peaks, filament-type saddle points, wall-type saddle points and voids. According to the sign of their eigenvalues, one can categorise these points into extrema having eigenvalues with identical sign (peaks with signature {\tiny ${-\!-\!-}$} and voids {\tiny ${+\!+\!+}$}) and saddle points having one eigenvalue with a different sign (filament-type saddles with signature {\tiny ${-\!-\!+}$} and wall-type saddles with signature {\tiny ${-\!+\!+}$}). We will also label the signature of a critical point by the number of negative eigenvalues, differing by one from three
for the peaks to zero for the voids.

\subsubsection{Finding critical points}
\label{sec:algo}
In order to identify critical points in a smoothed simulated field, a local quadratic estimation is used.
The detection relies on a second-order Taylor-expansion of the density field. About a critical point, this yields  
\begin{equation}
    \vvec{ x}-\vvec{x_c} \approx (\nabla\nabla \rho)^{-1} \nabla \rho ,
    \label{eq:finding_crit_point}
\end{equation}
where $\vvec{x_c}$ is the position of the critical point and $\rho$ is the density field.
The detection algorithm\footnote{The code is available at
\href{https://github.com/cphyc/py_extrema}{py-extrema}.}
then works as follows.
\emph{a)} For each cell in the grid, compute the gradient and Hessian of the density field,
\emph{b)} solve Equation~\eqref{eq:finding_crit_point} and discard all solutions found at a distance greater than one pixel ($\max_{i=1,\dots 3}(|\vvec{x}_i-\vvec{x}_c|) < 1\ \mathrm{pixel}$) and
\emph{c)} flag each cell containing a critical point. 
\emph{d)} Finally, loop over flagged cells that contain multiple critical points of the same kind, retaining for each only the critical point closest to the centre of the cell
\citep[see also][Appendix~G]{Gay2012}.
The numbers of identified peaks, filaments, walls and voids are tabulated in Table~\ref{tab:totnum}.
\begin{table}
	\centering
	\caption{Total number of peaks ($\mathcal{P}$), filaments ($\mathcal{F}$), walls ($\mathcal{W}$), and voids ($\mathcal{V}$) in different redshift bins.}
	\label{tab:totnum}	
    \begin{tabular}{cccccc} % six columns, alignment for each
        \hline\hline
        & $\mathcal{P}$ & $\mathcal{F}$ & $\mathcal{W}$ & $\mathcal{V}$\\
        \hline
		$0\le z<0.5$ & 3690 & 10542 & 10098 & 3209\\
		$0.5\le z<1$ & 3693 & 10712 & 10377 & 3324\\
		$1\le z<2$ & 3690 & 10857 & 10631 & 3434\\
		$2\le z<3$ & 3689 & 10999 & 10872 & 3542\\
		$3\le z<4$ & 3713 & 11138 & 11064 & 3620\\
		\hline
    \end{tabular}
\end{table}

\subsubsection{Computing  clustering properties}

To measure their clustering properties, we count the pairs of critical points with rarity above or below a certain threshold. The rarity of the critical point is defined as 
\begin{equation}
    \nu\equiv{\delta}/{\sigma},
\end{equation}
where $\delta$ is the over-density contrast of the smoothed density field
\begin{equation}
    \delta\equiv{\rho}/{\bar{\rho}}-1,
\end{equation}
and $\sigma$ is the rms fluctuation of the field
\begin{equation}
    \sigma^{2}\equiv\left\langle \delta^2\right\rangle.
\end{equation}
For peak and filament (resp. void and wall) critical points, we extract points with rarity higher (resp. lower) than a given threshold. In this paper, we 
fix this threshold as the rarity for each type of critical points, $\nu_{\rm type,c}$, yielding the same relative abundances (\eg 5, 10, 15, 20 \%) defined by the ratios
\begin{equation}
    \label{eq:rareness}
    \frac{N_{\rm type}(\nu \ge \nu_{\rm type,c})}{N_{\rm type}},%
\end{equation}
for peaks and filaments, and  
\begin{equation}
    \label{eq:rareness2}
    \frac{N_{\rm type}(\nu \le \nu_{\rm type,c})}{N_{\rm type}}, 
\end{equation}
for voids and walls, in each redshift bin.

Our purpose is to single-out the redshift evolution of the clustering properties by sampling the population that represents the same abundance for a given type of critical points. This somewhat \emph{ad hoc} choice allows us to limit the number of configurations we investigate, and imposes non-intersecting ranges of rarities ($\nu$) for peak and void on the one hand, and walls and filaments on the other. This choice was driven by symmetry, but has consequences: for instance, the chosen wall-saddle points do not represent well what one would identify as say, the great SDSS wall, which is an overdense feature of the LSS. The lack of overlap in rarities also impacts exclusion zones, see  Appendix~\ref{sec:cross-rare} for more details. Note however that all the machinery developed in this paper could easily be applied to any choice of abundances for the critical points. 

Using critical points extracted by applying these particular rarity thresholds, we measure the correlation functions with the estimator adopted in \citet{cadiou2020}, simply given by a Davis-Peebles estimator \citep{1983ApJ...267..465D}
\begin{equation}
1+\xi_{ij}(r) = \frac{\langle C_{i}C_{j}\rangle}{\sqrt{\langle C_{i}R_{j}\rangle \langle C_{j}R_{i}\rangle }}
\sqrt{\frac{N_{R_i} N_{R_j}}{N_{C_i} N_{C_j}}},
\label{eq:xi}    
\end{equation}
where $C_i$ denotes a particular catalogue of critical points $i\in\{{\cal P},{\cal F},{\cal W },{\cal V}\}$ and $R_i$ is a corresponding random catalogue with randomly distributed points following a uniform probability distribution in the same volume.  Here $\langle XY \rangle$ represents number counts of the pairs between $X$ and $Y$ whose separation is $r$. The sample size, $N_{R_i}$, of the random catalogue for extrema (resp. saddles) is a factor of $100$ (resp. 200) larger
than the correspondent size of our simulated datasets, $N_{C_i}$.
 Note that periodic boundary conditions are taken into account when computing  distances.

%%%%%%%%%%%%%%%%%%%%%%%%%%%%%%%%%%%%%%%%%%%%%%%%%%
\section{Counts and clustering measurements}\label{sec:measurements}
%%%%%%%%%%%%%%%%%%%%%%%%%%%%%%%%%%%%%%%%%%%%%%%%%%

Let us describe the results of our measurements in $\Lambda$CDM simulations. 
 The corresponding theory for Gaussian random fields is presented 
in Appendix~\ref{sec:theory}.
We will focus first on the total (Section~\ref{sec:Ncrit}) and differential number counts of critical points (Section~\ref{sec:oneptfunction})
as a function of rarity, then present their auto-correlations (Section~\ref{sec:auto}) before studying their cross-correlations (Section~\ref{sec:cross}). In particular, we will focus on the angular dependence of the cross-correlations involving saddle points as measured in their eigenframe (Section~\ref{sec:angle}) to understand how the geometry of the  cosmic web impacts clustering.
\subsection{Total number of critical points}
\label{sec:Ncrit}

As quoted in Table~\ref{tab:totnum}, we identified more peaks than voids. As expected, this deviates from Gaussian predictions for which those two numbers are equal \citep{BBKS}. Indeed, the non-linear evolution tends to break the symmetry between underdense and overdense regions and in particular between peaks and voids. The number difference between them is shown to be larger at lower redshifts. This is also true for filaments and walls which should appear in the same proportion in the Gaussian regime and should be approximately three\footnote{For GRF, the ratio between the numbers of filaments and peaks (or walls and voids) is exactly $(29\sqrt{15}+18\sqrt{10})/(29\sqrt{15}-18\sqrt{10})\approx 3.05$.} times more abundant than peaks and voids. \cite{Gay2012} showed that the first non-linear correction has the same amplitude for all kinds of critical points, with the same sign for peaks and filaments and the opposite sign for walls and voids. Hence, the total number of extrema (voids and peaks together) is preserved at first non-Gaussian order, similarly for the total number of saddles and therefore also their ratio. 
 Interestingly, this property pervades as the measured number ratio of saddles over extrema remains nearly constant (close to 3) in our simulations throughout the redshift range investigated.
Note that for sufficiently large volumes, we expect the ratio between the number of peaks and walls over filaments and voids to be very close to one throughout the redshift evolution, which is indeed the case here. This is because the topology of the box hence the genus which is the alternating sum of critical points should be preserved.

\subsection{One point function of critical points}
\label{sec:oneptfunction}

\begin{figure}
\centering	\includegraphics[trim=0 15bp 0 15bp,clip,width=1.1\columnwidth]{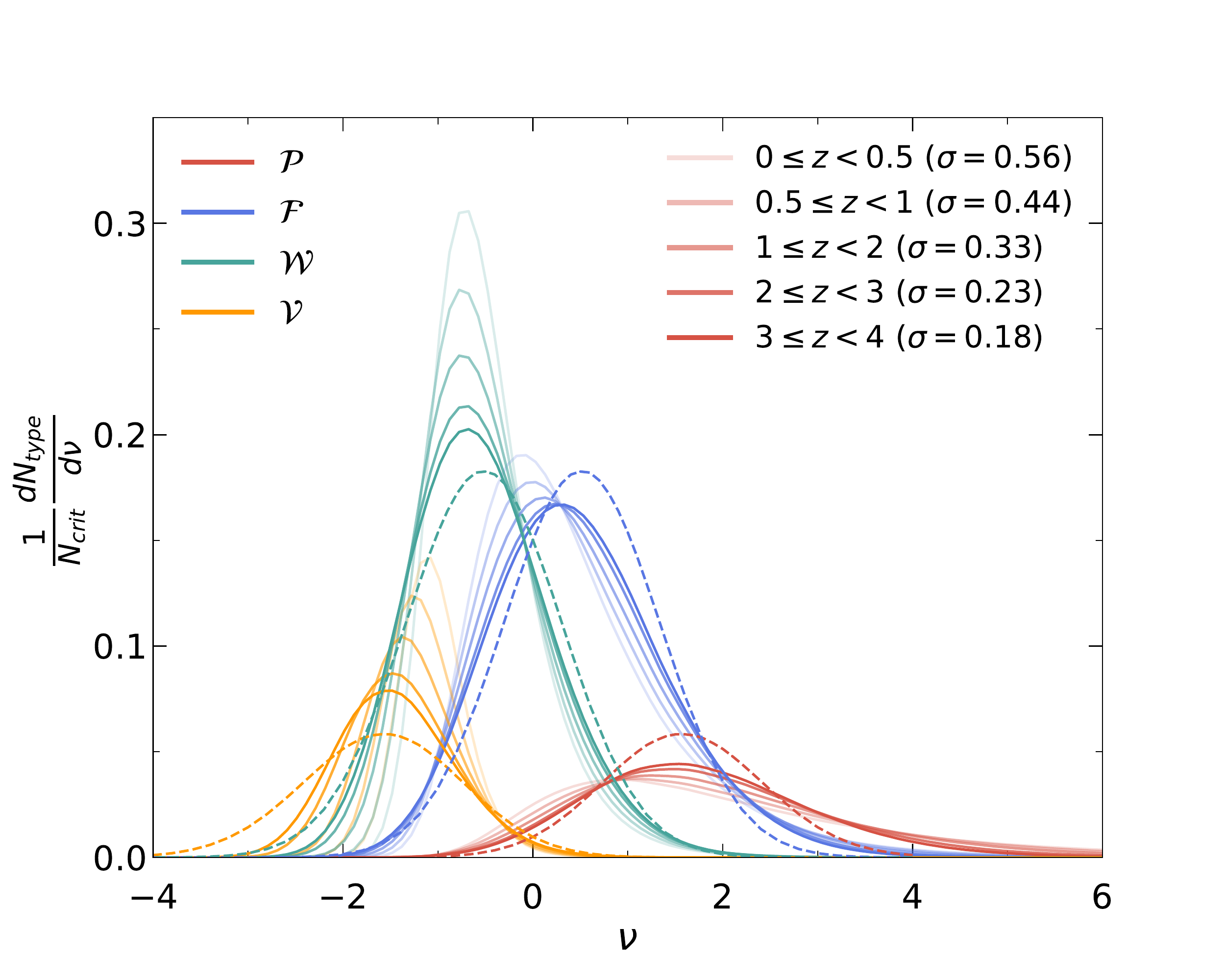}
    \caption{
    Relative number counts 
     of critical points (${\cal P}$: peaks, ${\cal F}$: filaments, ${\cal W}$: walls, and ${\cal V}$: voids) as functions of their rarity in different redshift bins and the corresponding  $\sigma(z)$ of the field, as labelled. The number counts in the Gaussian limit (dashed) are also displayed. At all redshifts, they show distinctive non-Gaussian asymmetry which becomes more pronounced at late time (as predicted by perturbation theory). 
    These counts will allow us to define thresholds containing a given percentage of all critical points of a given kind, when looking at auto and cross-correlations between critical points. Note how strikingly the maximum of the wall counts varies so little with redshift. The maximum amplitude of void and wall counts continuously increase from the Gaussian limit to the lowest redshift, while peak counts decrease. Interestingly, the maximum of the filament counts first decreases from the Gaussian prediction and then increases at later times. In most cases, the counts become narrower as they evolve, except for peaks.
    }
    \label{fig:histPDF}
\end{figure}
Let us now study the distribution of rarity of the measured critical points.
Figure~\ref{fig:histPDF} shows the relative number counts of critical points as functions of rarity. The number counts of a particular type is normalised by that of all critical points in each snapshot. These measurements underpin
the relations between rarity thresholds and abundances given in Table~\ref{tab:height_rareness}.
\begin{table*}
	\centering
	\caption{Threshold used on the rarities of critical points ($\nu_{\rm type,c}$) so as to keep a fixed abundance (given by the N$\%$ rarest objects) at various redshifts. These values are redshift-dependent because of the development of non-Gaussianities in the relative number counts (see Figure~\ref{fig:histPDF}). The last column also displays the predictions for Gaussian random fields.
    }
    
	\label{tab:height_rareness}	
    \begin{tabular}{cccccccc} % six columns, alignment for each
        \hline\hline
		type & abundance & $0\le z<0.5$ & $0.5\le z<1$& $1\le z<2$ & $2\le z<3$ & $3\le z<4$& $ z\rightarrow\infty$\\
		& & ($\sigma=0.56$) & ($\sigma=0.44$) & ($\sigma=0.33$) & ($\sigma=0.23$) & ($\sigma=0.18$)&  ($\sigma\rightarrow 0$)\\
		\hline
		$\mathcal{P}$ & $20\%$ & 3.43 &  3.28 & 3.08 & 2.85 & 2.73&2.30\\
		& $15\%$ & 3.98 &  3.73 & 3.45 & 3.15 & 2.99&2.46\\
		& $10\%$ & 4.74 &  4.36 & 3.95 & 3.55 & 3.33&2.67\\
		& $5\%$ & 6.09 &  5.42 & 4.77 & 4.18 & 3.87&2.98\\
		\hline
		
		$\mathcal{F}$& $20\%$ & 1.13 &  1.19 & 1.23 & 1.25 & 1.25&1.21\\
		& $15\%$ & 1.39 &  1.44 & 1.46 & 1.46 & 1.45&1.37\\
		& $10\%$ & 1.76 &  1.79 & 1.78 & 1.74 & 1.71&1.57\\ 
		& $5\%$ & 2.40 &  2.36 & 2.29 & 2.19 & 2.12&1.87\\
		\hline
		
		$\mathcal{W}$ & $20\%$ & -0.90 &  -0.97 & -1.04 & -1.10 & -1.14&-1.21\\
		& $15\%$ & -0.96 &  -1.05 & -1.13 & -1.21 & -1.26&-1.37\\
		& $10\%$ & -1.03 &  -1.14 & -1.24 & -1.35 & -1.41&-1.57\\
		& $5\%$ & -1.12 &  -1.26 & -1.39 & -1.53 & -1.62&-1.87\\
		\hline
		
		$\mathcal{V}$& $20\%$ & -1.24 &  -1.41 & -1.60 & -1.79 & -1.91&-2.30\\
		& $15\%$ & -1.27 &  -1.46 & -1.66 & -1.88 & -2.01&-2.46\\
		& $10\%$ & -1.31 &  -1.51 & -1.74 & -1.98 & -2.13&-2.67\\
		& $5\%$ & -1.37 &  -1.59 & -1.85 & -2.13 & -2.30&-2.98\\
		\hline
    \end{tabular}
\end{table*}
In the Gaussian limit, the distributions of peaks and voids (resp. filaments and walls) are symmetric
to each other about $\nu=0$ \citep{BBKS}. 
However, as the density field becomes non-linear and non-Gaussian, such symmetries break down \citep[see for instance][]{Gay2012}. At low redshift, we observe that the rarity distributions of critical points become narrower and have higher maximum amplitude except for peak critical points. 
The relative number counts for the peak and filament critical points clearly show a positively-skewed rarity distribution driven by gravitational clustering.
In particular, their relative number count at the highest end slightly increases with decreasing redshift as overdense regions become denser with time.
This behaviour is similar in essence to the so-called Edgeworth expansion
\begin{equation}
P(\nu)=\exp(-\nu^2/2)\left[1+\sigma S_3 H_3(\nu)+\dots \right] \label{eq:edge}
\end{equation}
of the density field (not restricted to critical points) whose mildly non-Gaussian PDF is well captured by perturbative techniques \citep{bernardeau&kofman95}, and even more so when a large-deviation principle is used to include consistently more non-linearities \citep{BCP13}. Similar techniques can be used in the context of critical points, as was for instance derived for arbitrary non Gaussianities in \cite{Gay2012}. Here, cumulants involving not only the density field -- as is the case for the skewness $S_3$-- but also its first and second derivatives appear. However, note that the prediction for the differential counts is not analytical (even at Gaussian order) but can be evaluated numerically. 

Qualitatively, the behaviour of the redshift evolution in Figure~\ref{fig:histPDF} seems consistent with \eg Figure~6 in \cite{Gay2012} (which in turn was shown to agree with perturbation theory predictions), and allows us to probe the range of non-linearity $\sigma > 0.18$ that starts where \cite{Gay2012} analysis ends, $\sigma < 0.2$. 
In particular, the evolution from the initial Gaussian conditions to the highest redshift range having similar $\sigma$ shows a very good agreement. For example, the position of the maximum of the distribution displays almost no evolution for peaks and voids but shifts to the left for walls and filaments. The maximum becomes higher compared to the Gaussian prediction for voids and walls, whereas it decays for filaments and peaks. Interestingly, the trend of the maximum moving toward rarities of smaller magnitude persists to the lowest redshift bins for all critical points except walls. For walls, the position of the maximum is almost stationary at $z<3$ after having moved to the left from the initial maximum position in Gaussian limit. On the other hand, the height evolution of the maximum is monotonic for all types except for filaments, for which the evolution happens in two successive stages. As seen in \cite{Gay2012} (and tentatively in the very high redshift bins here), the maximum of the relative number count first goes down and then we find it to increase again when non-gaussianities continue to rise.

From the differential number counts of critical points, we can identify the thresholds that will be used to compute correlation functions. Table~\ref{tab:height_rareness} provides the mean thresholds $\nu_{\rm type,c}$ for all types of critical points and various abundances in the five redshift bins we consider. As expected, the threshold increases for rarer peaks as the field becomes more non-Gaussian with redshift.
Notable is a relative stability with redshift of the rarity threshold for filament-type saddle points
at fixed abundance except for the rarest abundance threshold for which an increase with non-linearity is observed similar to peaks. We also note the trend for underdense voids and walls opposite to peaks, for which the rarity threshold decreases in magnitude as the field evolves, probably as a result of the positive skewness that dominates in the first stage of structure formation.

%%%%%%%%%%%%%%%%%%%%%%%%%%%%%%%%%%%%%%%%%%%%%%%%
\subsection{Auto-correlation of critical points}
\label{sec:auto}
Having characterised the one-point statistics of critical points, let us now turn to their auto-correlations.

\begin{figure*}
	\includegraphics[trim=0 120bp 0 120bp,clip,width=2\columnwidth]{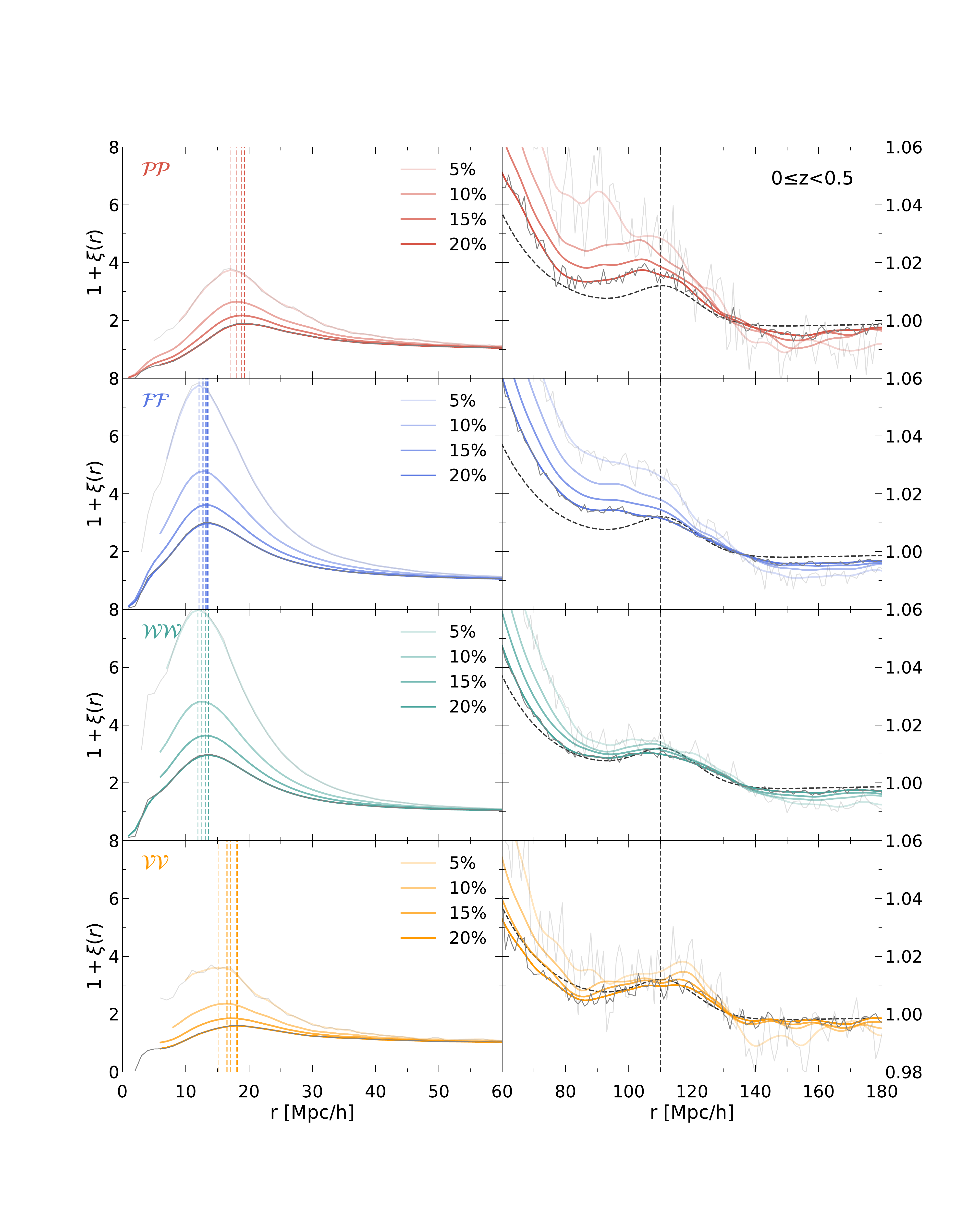}
    \caption{Auto-correlations of critical points with different abundances for separations between 0 and 60\mpcph\ ({\it left}) and BAO scales ({\it right})  for $0\le z<0.5$. $\cal PP$ (peak-peak), $\cal FF$ (filament-filament), $\cal WW$ (wall-wall), and $\cal VV$ (void-void) correlations are shown in panels from top to bottom. Smoothed versions (non-grey) of the measurements and unsmoothed measurements (grey) for abundances between $5\%$ and $20\%$ are presented. The linear matter correlation function at the corresponding median redshift \citep[multiplied by $8$ which is close to the square of the bias factor 
    for $\nu_{\rm peak} \ge2.5$ for peak critical points, ][]{matsubara&codis19} is also displayed on the right-hand panels
    with a dashed line. The vertical lines mark resp. the position of the maximum of the auto-correlations ({\it left}) and the BAO peak position of the linear matter correlation ({\it right}). These positions are found as the local maximum using cubic-spline interpolation. Note that for $\cal FF$ and $\cal WW$ they follow closely each other. For rarer critical points, the maxima of the correlations move to the left and become higher. At late time, the BAO positions are not very well defined regardless of the abundance in most cases. Finally note that the inflection point at $\sim 133$ \mpcph\ corresponds to zero correlation for all cases.
    } 
    \label{fig:auto_h}
\end{figure*}

\subsubsection{Rarity dependence of the auto-correlations: Figure~\ref{fig:auto_h}}
Let us first study the auto-correlation of critical points at low redshift for different rarities, before studying their redshift evolution.
Figure~\ref{fig:auto_h} shows the auto-correlations of critical points with different abundances between 5 and $20\%$ and for low redshifts, $0\le z<0.5$. We split the plots for separations between 0 and 60\mpcph\ (left panels) and BAO scales (right panels). 

Let us recall that, for peaks and filaments (resp. voids and walls), we extract critical points above (resp. below) the rarity yielding a particular abundance (see Table~\ref{tab:height_rareness}).

The shape of the auto-correlation functions always follows the same qualitative pattern: at small separations, they are negative corresponding to a region of anti-clustering ($\xi < 0$), then increase and get positive reaching a maximum at separations $\approx 3 R_{\rm G}$
for peaks and voids and $\approx 2.2 R_\mathrm{G}$ for filaments and walls, before gradually decreasing towards zero in the large-separation regime where bias expansions can be performed \citep{Desjacques2018}. 
This behaviour has been investigated in details for Gaussian peaks in many works including in 1D \citep{baldauf+16} and in 3D (Baldauf et al., in prep.) where in particular the typical size of 

the anti-clustering region
and its scaling with the peak heights difference $\propto (\Delta \nu)^{1/3}$ were derived. Note that if one naively associates
Gaussian $\mathrm{FWHM}=2.35\; R_\mathrm{G}$ with a size of peak  footprint, finding the correlation maximum at $1.3-1.4\; \mathrm{FWHM}$ shows that rare peaks tend to come in rather tightly packed clusters.

Let us now describe our findings for the dependence of the auto-correlations with the rarity 
of the objects, based on Figure~\ref{fig:auto_h}.
At a general level, as shown in the left panels, the main trends found for all types of critical points with increasing rarity are threefold: enhanced clustering (\ie larger overprobability $1+\xi$), a maximum for the 2pCF occurring at a radius that is only weakly decreasing,   and a narrower anti-clustering region.

In more detail, at large separations $ r \gtrsim 5\; R_\mathrm{G} \approx 30 h^{-1} \mathrm{Mpc}$, the increase of the auto-correlation amplitude for rarer critical points, implying a stronger clustering, 
can be explained with a simple linear bias\footnote{$b_{10}(\nu)\sim \nu/\sigma_0$ in the high $\nu$ limit, see Appendix~\ref{sec:theory}.} of critical points relative to the dark matter correlation function $\xi_{0,0}$, $\xi_{\cal PP}\sim b_{10}^2(\nu)\xi_{0,0}/\sigma_0^2\sim \nu^2 \xi_{0,0}/\sigma_0^2$ which reflects a known effect of stronger clustering of more biased tracers \citep{Kaiser1984, desjacques08,Uhlemann:2016un,Desjacques2018}.
One can go beyond the simple linear bias and derive a peak bias expansion assuming all correlation functions are small compared to their zero lag (see also Appendix~\ref{sec:theory}). We refer the reader to the review \cite{Desjacques2018} for more details. Predictions up to order three \citep{matsubara&codis19} have been obtained for Gaussian peaks but the convergence of these expansions is slow and cannot capture the small and intermediate scales including the maximum of the 2pCF and even more so their exclusion zone for which a full numerical integration is required. They are nonetheless very useful to understand the large separation region including BAO scales as was extensively studied by notably \cite{desjacques08, 2010PhRvD..82j3529D,2017PhRvD..95d3535B}. 

In the intermediate regime of the 2pCF the main feature is
the maximum of the auto-correlation. While its height continues to reflect the larger bias and more pronounced clustering of the rarer objects, we also point out that
the position of the maximum correlation shifts toward lower
separations as rarity increases (at least for peaks and voids), an
approximately ten percent effect between $5\%$ and $20\%$ abundant peaks at $0\le z<0.5$.
Let us also note that the height of the maximum of the auto-correlations for filaments and walls are about a factor of two higher than those for peaks and voids for a given abundance. This may be related to different curvature conditions, via the second derivatives of the density field, between extrema and saddles. Note also that the radius of maximum correlation is typically smaller for saddles than for extrema.
Quantities at special locations including the maximum at intermediate separation and inflection point at large scale of the correlation functions for various abundances are listed in Table~\ref{tab:peak_posh_rarity}. We calculate the standard deviation of the mean quantities at the lowest redshift snapshot in each redshift bin as a proxy to the typical scatter of the mean quantities in each redshift bin. Indeed, since measurements in consecutive snapshots are correlated, measuring  errors in the full bin would likely under-estimate them.

The small shift of the maximum of the 2pCF towards the left with rarer critical points can be explained with rarity difference ($\Delta \nu$) that is usually smaller in rarer critical point samples. Indeed, the extent of the anti-clustering region decreases with decreasing rarity difference \protect{\citep{baldauf+16}} and hence the position of the maximum correlation moves to a smaller radius. For Gaussian random fields with the same power spectrum, the mean rarity difference can be computed and yields $\overline{\Delta \nu}=\{0.43,0.40,0.37,0.33\}$ for abundances 20, 15, 10 and $5\%$ respectively. If the shift of the maximum scales like $\Delta\nu^{1/3}$, as derived by Baldauf+ in prep., then we expect a shift of 0, 2, 5 and $8 \%$ of $r_{\rm max}$ with respect to the 20$\%$ abundance case. If the redshift evolution is small (as will indeed be shown in Section~\ref{sec:evolution-auto}), then those numbers can be compared to the relative shifts measured in the simulation as reported in Table~\ref{tab:peak_posh_rarity} namely $\{0\pm 2\%,3\pm  2\%,7\pm 3\%,11\pm 2\%\}$ and seem perfectly consistent.

\begin{table*}
	\centering
	\caption{Position and height of the maximum as well as the location of the inflection point in auto- and cross- correlations as a function of rarity for the lowest redshift bin, $0\le z <0.5$. The errors are the standard deviations of the mean obtained from the resampled realisations at $z=0$ using bootstrapping.
	}
	\label{tab:peak_posh_rarity}	
    \begin{tabular}{ccccc} % six columns, alignment for each
        \hline\hline
		type & abundance & $r_{\rm max}$ [\mpcph] & $h_{\rm max}$ & $r_{\rm inf}$ [\mpcph]\\
		\hline
		$\mathcal{PP}$ & $20\%$ & $19.3\pm0.30$ & $1.9\pm0.01$ & $132.7\pm1.8$\\
		& $15\%$ & $18.8\pm0.30$ & $2.2\pm0.02$ & $131.9\pm2.1$\\
		& $10\%$ & $18.0\pm0.55$ & $2.7\pm0.03$ & $135.7\pm2.4$\\
		& $5\%$ & $17.1\pm0.48$ & $3.7\pm0.07$ & $133.7\pm3.9$\\
		\hline
		
		$\mathcal{FF}$ & $20\%$ & $13.5\pm0.08$ & $3.0\pm0.01$ & $134.3\pm1.5$\\
		& $15\%$ & $13.2\pm0.11$ & $3.6\pm0.01$ & $133.6\pm1.5$\\
		& $10\%$ & $12.7\pm0.15$ & $4.8\pm0.02$ & $134.4\pm1.7$\\
		& $5\%$ & $12.1\pm0.14$ & $7.8\pm0.06$ & $133.3\pm1.6$\\
		\hline
		
		$\mathcal{WW}$ & $20\%$ & $13.6\pm0.12$ & $3.0\pm0.01$ & $136.0\pm1.2$\\
		& $15\%$ & $13.1\pm0.14$ & $3.6\pm0.01$ & $134.3\pm1.3$\\
		& $10\%$ & $12.5\pm0.14$ & $4.8\pm0.03$ & $134.5\pm1.6$\\
		& $5\%$ & $11.9\pm0.17$ & $8.0\pm0.07$ & $134.5\pm3.0$\\
		\hline
		
		$\mathcal{VV}$ & $20\%$ & $18.1\pm0.30$ & $1.6\pm0.01$ & $132.3\pm1.6$\\
		& $15\%$ & $17.1\pm0.76$ & $1.9\pm0.02$ & $131.5\pm1.2$\\
		& $10\%$ & $16.5\pm1.12$ & $2.4\pm0.03$ & $132.0\pm2.2$\\
		& $5\%$ & $15.2\pm0.96$ & $3.6\pm0.10$ & $131.8\pm1.9$\\
		\hline\hline
		
		$\mathcal{PF}$ & $20\%$ & $13.6\pm0.05$ & $4.9\pm0.01$ & $133.2\pm1.1$\\
		& $15\%$ & $13.7\pm0.05$ & $6.1\pm0.02$ & $133.5\pm1.1$\\
		& $10\%$ & $13.9\pm0.05$ & $8.2\pm0.03$ & $134.5\pm1.3$\\
		& $5\%$ & $13.9\pm0.06$ & $13.5\pm0.08$ & $132.9\pm1.7$\\
		\hline
		
		$\mathcal{WV}$ & $20\%$ & $14.6\pm0.06$ & $3.8\pm0.01$ & $132.2\pm2.4$ \\
		& $15\%$ & $14.7\pm0.07$ & $4.5\pm0.02$ & $132.7\pm2.2$\\
		& $10\%$ & $14.8\pm0.08$ & $5.8\pm0.03$ & $133.6\pm2.9$\\
		& $5\%$ & $15.0\pm0.09$ & $8.8\pm0.06$ & $132.3\pm3.1$\\
		\hline
    \end{tabular}
\end{table*}

Getting to even smaller scales, all critical point auto-correlation functions seem to exhibit a
zone of negative correlations, as was found before for peaks \citep{baldauf+16,codis+18}. From theoretical considerations, the limit $\xi(0)=-1$ is expected, though the resolution effects of our measurements do not allow to probe
the immediate vicinity of the origin.
The suppression of the probability of finding another critical points next to the first one happens because of constraints on curvature and height in the smoothed correlated density field.
For instance, it is very unlikely to find two peaks infinitely close to each other unless the peaks have the exact same height \citep{baldauf+16}.
One can notice that the crossing of the Poisson case $\xi=0$ seems to shift closer to the origin for larger rarity (extrapolating our measurements in the low separation regime).

Overall, the three main features at small separation are consistent with the prediction for Gaussian random fields (see Figure~\ref{fig:autoGRF} in Appendix~\ref{sec:GRF}) where we find similar rarity dependencies.
In particular, as in the measurements, the Gaussian prediction also shows that filament correlations have amplitudes approximately twice as large as for peaks or
that positions of the correlation maxima are somewhat closer to the origin for rarer points, in agreement with the estimate from Baldauf+ in prep. as explained above.
It is notable how well the rarity dependencies of the auto-correlations of the critical points in the relatively non-linear regime ($\sigma \approx 0.6$) are already encoded in the initial perturbations. The evolution of auto-correlations of critical points thus preserve the initial clustering properties, supporting the ``Cosmic Web'' paradigm of \cite{bkp96}.

Let us now focus on the right panels of Figure~\ref{fig:auto_h} which show the auto-correlations at very large separations where correlations exhibit the BAO effect. To qualitatively describe how the large-scale features of the clustering of critical points differ from that of the dark matter, we also compute the (unsmoothed) linear dark matter correlation function by adopting the transfer functions in \citet{eisenstein&hu98} \footnote{The smoothing typically reduces the sharpness of the BAO feature and may slightly shift the peak position inward.}. We multiply the linear correlation by a factor of $8$ so that it has a similar amplitude to the measured correlations of the 20\% rarest peak critical points. This factor can be qualitatively interpreted as the square of the large-scale bias, and reflects  that the correlation is enhanced compared to the linear prediction \citep{Kaiser1984}. Our choice of bias factor for the 20\% rarest peaks (with $\nu_{\rm peak}>3.5$, see Table~\ref{tab:height_rareness}) is qualitatively consistent with the result obtained in \citet{matsubara&codis19}, in which their bias factor is approximately $2.6$ for peaks with $\nu_{\rm peak}>2.5$, using a perturbative approach in the large-separation limit.

In general, all auto-correlations (except for  filaments) decrease from $r \approx 20$ to $80$\mpcph.
Then, they gently increase at around $100\le r<120$\mpcph\ showing BAO-like features before dropping again at larger separation. However,  the exact position of the BAO bumps are not robustly defined for all abundances: further simulations  will be required to make a more  quantitative comparison because the correlation function is too noisy at late time.
As expected for biased tracers in general, the BAO wiggles are found to be enhanced compared to the dark matter case, an effect which persists despite the smoothing and the non-linear gravitational evolution which are known to reduce the sharpness of the BAOs. Around the BAO peak, we also find that peak and filament auto-correlations vary more with abundance than voids and walls, which might be due to very overdense regions being more prone to non-linear effects including large-scale velocity flows, that also tend to shift the BAO peak to smaller scales.
Interestingly, correlation is larger for rarer points below $\approx 133$\mpcph\, but it is more anti-correlated at larger radii for all types of critical points. At roughly $r\approx 133$\mpcph, the correlation goes through zero through an inflection point (corresponding to $\xi(r_{\rm inf})=0$ and where correlations for different abundances almost coincide) for all abundances, a feature which is also seen in the linear matter correlation. It is remarkable that most correlations go through this inflection point at the same scale (see Table~\ref{tab:peak_posh_rarity}). 
It is surprisingly consistent with a multiplicative linear bias model at that scale, see Appendix~\ref{sec:theory} and would be interesting to study further as a possible robust cosmological probe (less sensitive to non-linearities). This situation seems similar to the so-called Linear Point \citep{Anselmi_2015,O_Dwyer_2020} which is the mid-point between the first dip and peak of the BAO pattern at $\sim 100$\mpcph\,. This point in the halo correlation function was shown to be robust against non-linear evolution, biasing effects and redshift space distortions and eventually can serve as a purely geometric cosmological test without the need for assuming any prior on the cosmological model or any model for the late-time galaxy correlation function evolution. Similar features seem to occur here for this secondary inflection point in the critical point 2pCFs and would therefore be worth investigating in more details in the future.

\begin{figure*}
	\includegraphics[trim=0 120bp 0 120bp,clip,width=2\columnwidth]{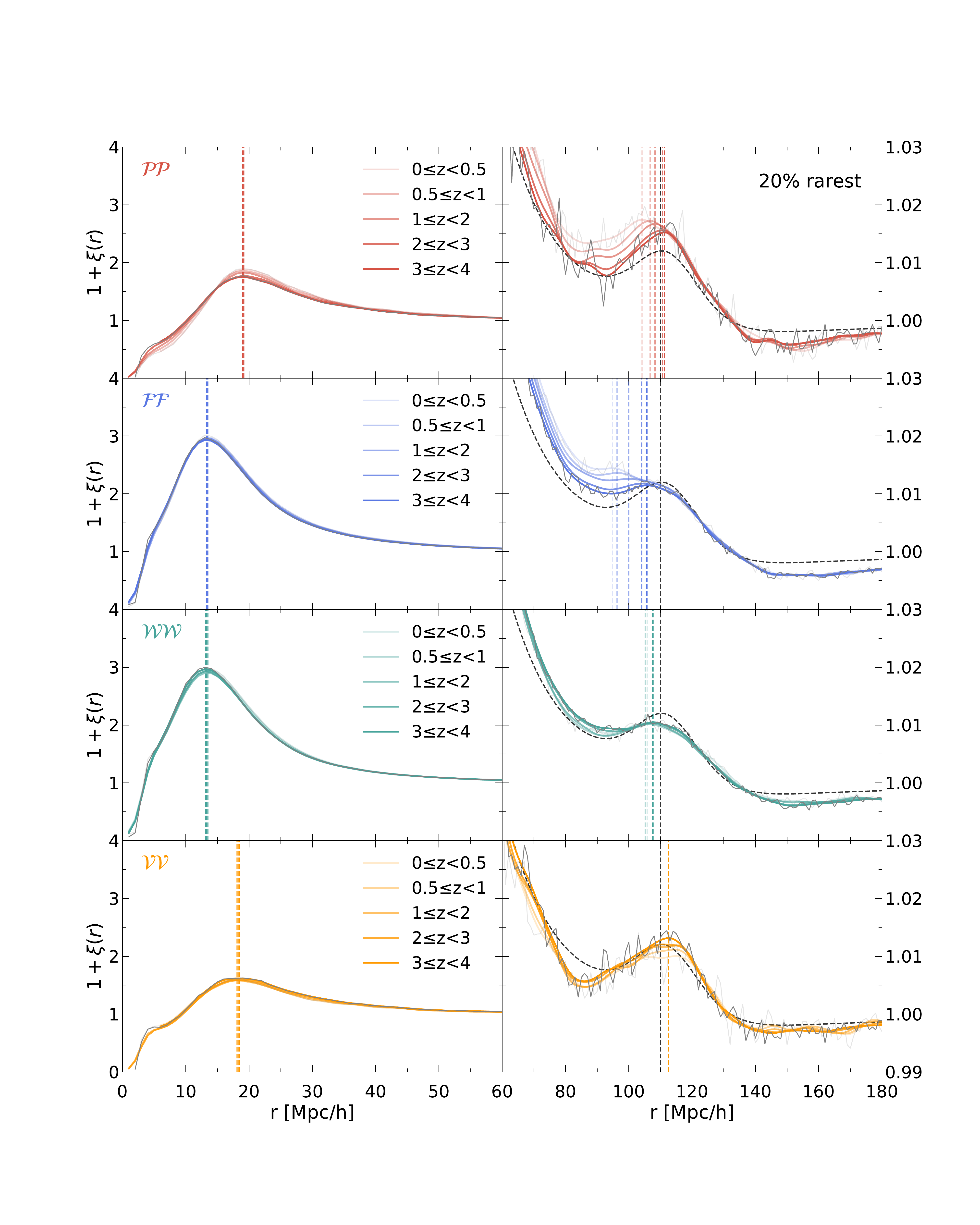}
    \caption{Auto-correlations of $20\%$ rarest critical points in different redshift bins for separations between 0 and 60 \mpcph\  ({\it left}) and BAO scales ({\it right}), following the same convention for the correlation type as in Figure~\ref{fig:auto_h}. We present both smoothed (coloured) and unsmoothed (grey) versions of the measurements for the lowest and highest redshift bins only. The linear dark matter correlation function (dashed) at the median redshift of $0\le z <0.5$ (multiplied by a factor of $8$) is also presented. The vertical lines mark the positions of the maximum correlation ({\it left}) and BAOs ({\it right}) of the auto-correlation (non-black) and the linear matter correlation (black). These positions are found as the local maximum using cubic-spline interpolation. Positions for the only well-defined BAO peaks are marked at large separation. The position and height of the maximum correlation barely change with redshift. Again, $\cal FF$ and $\cal WW$ correlations follow each other, and $\cal PP$ and $\cal VV$ also are similar although $\cal PP$ case has slightly higher maximum. The BAO signal is clearly detected and amplified compared to the linear matter correlation and it is more enhanced for extrema than for saddles. Note that the BAO position is almost stationary in $\cal WW$ correlations.
    } 
    \label{fig:auto_z}
\end{figure*}

\subsubsection{Redshift evolution of the auto-correlations: Figure~\ref{fig:auto_z}} \label{sec:evolution-auto}

Let us now turn to the redshift evolution of the auto-correlations of critical points in Figure~\ref{fig:auto_z}. For a quantitative comparison, we provide the positions and heights of the maximum and the BAO as a function of redshift in Table~\ref{tab:peak_bao_redshift}. At small and intermediate separations ($r\le30$\mpcph), the shape and height of the auto-correlations show little redshift evolution. In particular, the position change of the maximum correlation is at most $\Delta r \sim0.4$\mpcph\ for $0\le z<4$, which is 
small compared to the smoothing scale ($R_{\rm G}=6$\mpcph).

\begin{table*}
	\centering
	\caption{Positions and heights of the maximum and BAO peak in auto- and cross- correlations as a function of redshift for $20\%$ abundance. The errors are the standard deviations of the mean obtained from the resampled realisations using a bootstrap method at the lowest redshift snapshot of each redshift bin. Less well-defined BAO quantities are omitted.
	}
	\label{tab:peak_bao_redshift}	
    \begin{tabular}{cccccc} % six columns, alignment for each
        \hline\hline
		type & redshift & $r_{\rm max}$ [\mpcph] & $h_{\rm max}$ & $r_{\rm BAO}$ [\mpcph] & $h_{\rm BAO}$\\
		\hline
		$\mathcal{P\!P}$
		& $0\le z<0.5$ & $19.3\pm0.30$ & $1.88\pm0.01$ & $104.2\pm2.4$ & $1.017\pm0.001$\\
		& $0.5\le z<1$ & $19.0\pm0.43$ & $1.86\pm0.01$ & $106.7\pm1.8$ & $1.017\pm0.001$\\
		& $1\le z<2$ & $19.1\pm0.23$ & $1.83\pm0.01$ & $108.3\pm1.2$ & $1.017\pm0.001$\\
		& $2\le z<3$ & $19.1\pm0.32$ & $1.77\pm0.01$ & $110.6\pm1.9$ & $1.016\pm0.001$ \\
		& $3\le z<4$ & $19.0\pm0.29$ & $1.75\pm0.01$ & $111.3\pm0.9$ & $1.016\pm0.001$\\
		\hline
		
		$\mathcal{F\!F}$
		& $0\le z<0.5$ & $13.5\pm0.08$ & $2.98\pm0.01$ & $94.8\pm2.5$ & $1.014\pm0.001$\\
		& $0.5\le z<1$ & $13.5\pm0.10$ & $2.97\pm0.01$ & $96.3\pm1.5$ & $1.014\pm0.001$\\
		& $1\le z<2$ & $13.4\pm0.08$ & $2.96\pm0.01$ & $100.0\pm2.3$ & $1.013\pm0.001$\\
		& $2\le z<3$ & $13.4\pm0.08$ & $2.95\pm0.01$ & $104.1\pm2.4$ & $1.012\pm0.001$ \\
		& $3\le z<4$ & $13.3\pm0.07$ & $2.93\pm0.01$ & $105.7\pm1.2$ & $1.012\pm0.001$\\
		\hline
		
		$\mathcal{W\!W}$
		& $0\le z<0.5$ & $13.6\pm0.12$ & $2.96\pm0.01$ & $105.8\pm1.7$ & $1.010\pm0.001$\\
		& $0.5\le z<1$ & $13.5\pm0.11$ & $2.94\pm0.01$ & $105.2\pm1.2$ & $1.010\pm0.001$\\
		& $1\le z<2$ & $13.4\pm0.10$ & $2.91\pm0.01$ & $107.4\pm1.0$ & $1.010\pm0.001$\\
		& $2\le z<3$ & $13.2\pm0.12$ & $2.93\pm0.01$ & $107.7\pm1.0$ & $1.011\pm0.001$ \\
		& $3\le z<4$ & $13.2\pm0.08$ & $2.97\pm0.01$ & $107.4\pm2.3$ & $1.010\pm0.001$\\
		\hline
		
		$\mathcal{V\!V}$
		& $0\le z<0.5$ & $18.1\pm0.30$ & $1.60\pm0.01$ & - & - \\
		& $0.5\le z<1$ & $18.0\pm0.30$ & $1.58\pm0.01$ & - & - \\
		& $1\le z<2$ & $18.2\pm0.50$ & $1.58\pm0.01$ & - & - \\
		& $2\le z<3$ & $18.3\pm0.54$ & $1.59\pm0.01$ & - & - \\
		& $3\le z<4$ & $18.5\pm0.90$ & $1.62\pm0.01$ & $112.6\pm5.8$ & $1.013\pm0.001$\\
		\hline\hline
		
		$\mathcal{P\!F}$
		& $0\le z<0.5$ & $13.6\pm0.05$ & $4.93\pm0.01$ & $96.4\pm3.3$ & $1.015\pm0.001$\\
		& $0.5\le z<1$ & $13.6\pm0.05$ & $4.76\pm0.01$ & $96.9\pm5.2$ & $1.014\pm0.001$\\
		& $1\le z<2$ & $13.7\pm0.05$ & $4.58\pm0.01$ & $106.2\pm4.7$ & $1.013\pm0.001$\\
		& $2\le z<3$ & $13.8\pm0.05$ & $4.43\pm0.01$ & $108.0\pm2.4$ & $1.013\pm0.001$ \\
		& $3\le z<4$ & $13.7\pm0.04$ & $4.34\pm0.01$ & $108.1\pm1.2$ & $1.013\pm0.001$\\
		\hline
		
		$\mathcal{W\!V}$
		& $0\le z<0.5$ & $14.6\pm0.06$ & $3.78\pm0.01$ & $105.9\pm3.0$ & $1.010\pm0.001$\\
		& $0.5\le z<1$ & $14.5\pm0.06$ & $3.82\pm0.01$ & $106.0\pm3.9$ & $1.010\pm0.001$\\
		& $1\le z<2$ & $14.4\pm0.06$ & $3.84\pm0.01$ & $109.8\pm3.1$ & $1.011\pm0.001$\\
		& $2\le z<3$ & $14.2\pm0.06$ & $3.92\pm0.01$ & $110.2\pm2.5$ & $1.011\pm0.001$ \\
		& $3\le z<4$ & $14.1\pm0.05$ & $4.01\pm0.01$ & $109.1\pm1.3$ & $1.011\pm0.001$\\
		\hline
    \end{tabular}
\end{table*}

On BAO scales, wiggles are clearly amplified compared to the underlying dark matter field, the features appear particularly sharp for peaks and voids and at high redshift. We detect a redshift evolution of the peak and filament auto-correlations on the left-hand side of the BAO peak. In particular, the position and height of the BAO bump change non-negligibly. For these overdense critical points, the height of the bump increases at lower redshift and its location moves toward smaller scale. Also, the bumps become less pronounced as they evolve. This is again probably because the non-linear gravitational evolution of  over-dense regions smears out the BAO signature through pairwise motions \citep{desjacques08}. On the other hand, the auto-correlations of walls and voids show very little redshift evolution, which could be a good reason for using these underdense regions as cosmological probes. Noticeably, the position and height of the BAO bump is most stable for walls at $r\approx106-108$\mpcph. Since the two-point statistics of walls is almost unchanged from high to low redshift, this is in line with the findings by \citet{Gay2012} that one-point statistics of saddle critical points, specifically, 
can be modelled analytically down to the mildly non-linear regime.

%%%%%%%%%%%%%%%%%%%%%%%%%%%%%%%%%%%%%%%%%%%%%%%%%%%%%%%%%%%%%%%%%%%%%%%%%%%%%%%%%%%%%%%%%%
\subsection{Cross-correlation of critical points}
\label{sec:cross}

Let us now quantify the evolution of the cross-correlations of critical points as a function of rarity and redshift.

\begin{figure*}
	\includegraphics[trim=0 120bp 0 120bp,clip,width=2\columnwidth]{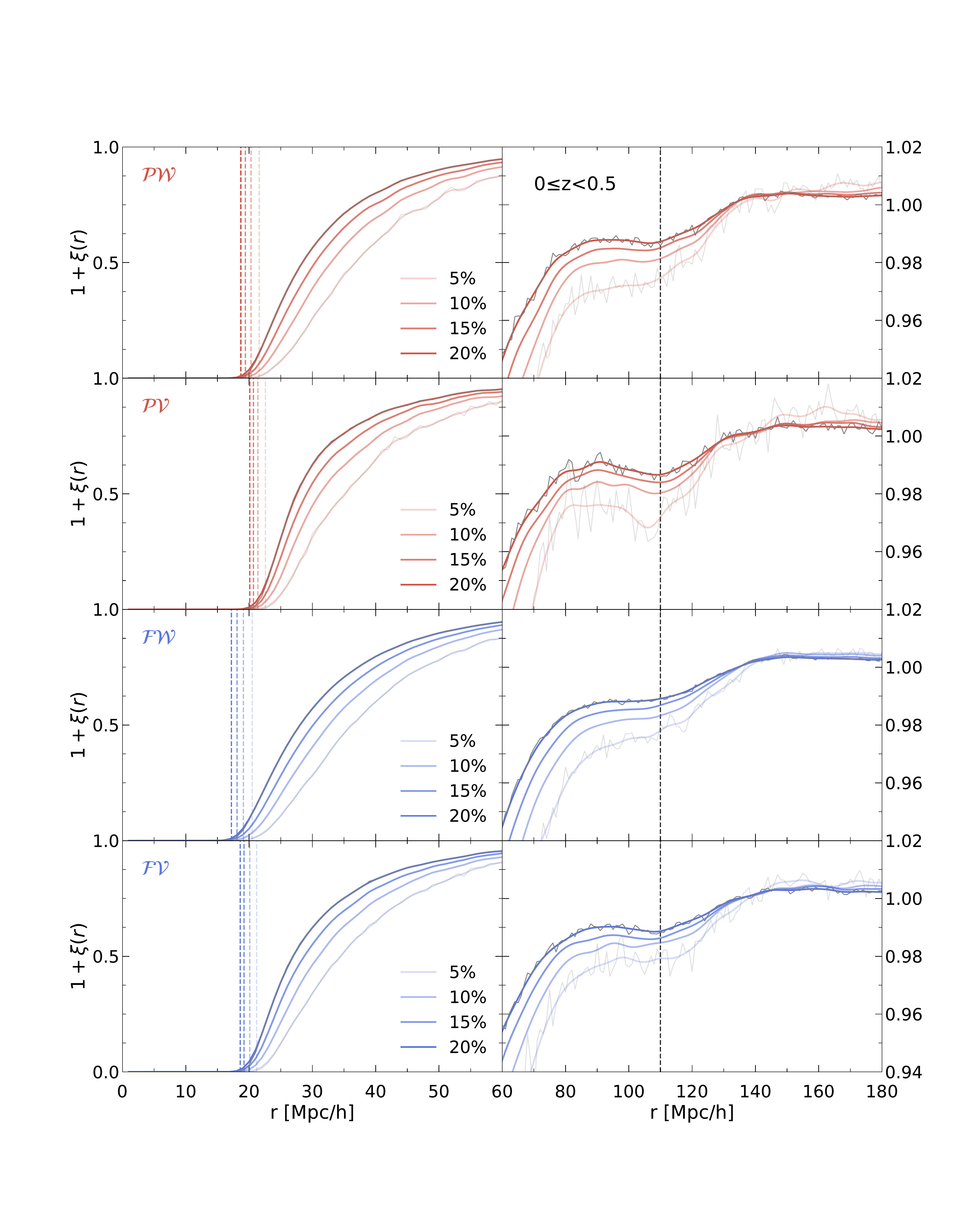}
    \caption{Cross-correlations of critical points with different abundance for separations between 0 and 60\mpcph\ ({\it left}) and BAO scales ({\it right}) and for $0\le z <0.5$. ${\cal PW}$ (peak-wall), ${\cal PV}$ (peak-void), ${\cal FW}$ (filament-wall), and ${\cal FV}$ (filament-void) correlations are shown in panels from top to bottom. For these cross-correlations, the vertical lines mark the size of exclusion zone ({\it left}) and the BAO position of the linear matter correlation ({\it right}). In contrast to the auto-correlations, the exclusion zone is well defined, and its size changes with abundance (as marked by the vertical lines). Because of the curvature constraints at critical points, the exclusion zone increases with the number of  eigenvalues with different sign between two critical points. Interestingly, at both small and large separation, the impact of abundance is significant. The BAO signature now corresponds to a dip in the correlation  rather than a bump,
    which, given our choice of rarity, is mainly driven by the fact that one out of two bias factors is negative. 
    }
    \label{fig:cross1_h}
\end{figure*}

\begin{figure*}
	\includegraphics[trim= 0 30bp 0 30bp,clip,width=2\columnwidth]{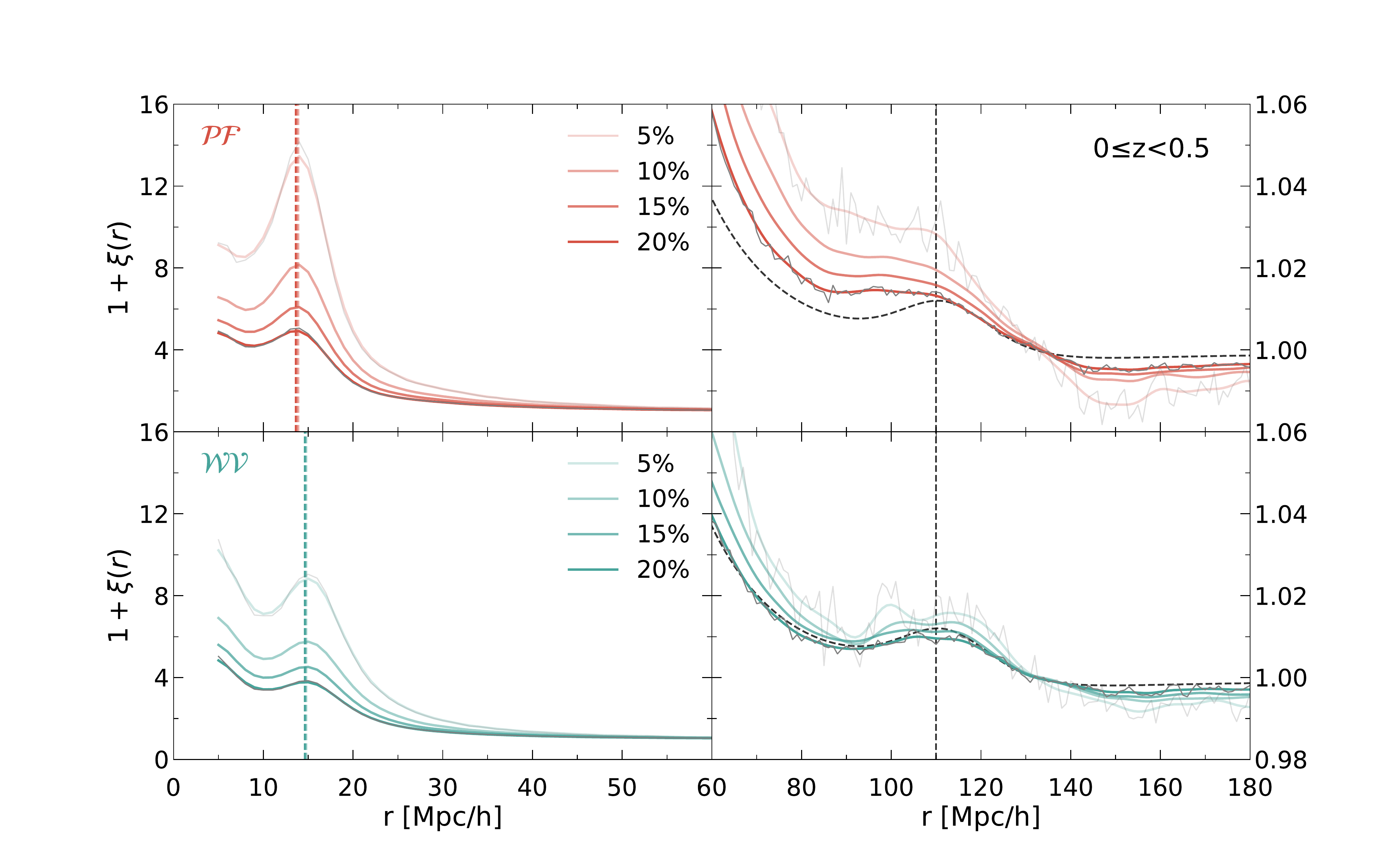}
    \caption{Same as Figure~\ref{fig:cross1_h}, but for ${\cal P\!F}$ (peak-filament) and ${\cal W\!V}$ (wall-void) correlations in the  top and bottom panels, respectively. For these cross-correlations, the vertical lines mark the positions of the maximum ({\it left}) and the BAO feature ({\it right}). We truncate the correlation function below $r<5$\mpcph{}, since the pair counts at separation near the resolution scale ($\sim2$\mpcph) are numerically suppressed.
    The linear dark matter correlation function (dashed) at the median redshift of $0\le z <0.5$ (multiplied by a factor of 8) is also presented. Strikingly, these cross-correlations have a maximum plus a divergence-like shape at small separation, reflecting a shell-like structure in the distribution of saddles around a given extremum (see Figure~\ref{fig:PFevec}). In fact, the ${\cal P\!F}$ correlation of Gaussian random field also diverges at very small separation showing a head and shoulder pattern (see Figure~\ref{fig:pf-abundancesGRF} and also Fig.~4 in \citet{codis+18}). Note that the amplitude of the  maximum in the left panel is higher for overdense pairs, \ie $\cal P\!F$. A clear abundance dependence is shown at small separation but it is less pronounced in the $\cal W\!V$ case at large separation. The BAO bumps shown in the cross-correlations are broader than that in the linear dark matter correlation.
    }
    \label{fig:cross2_h}
\end{figure*}

\subsubsection{Rarity dependence of  cross-correlations: Figures~\ref{fig:cross1_h} and~\ref{fig:cross2_h}}

Figure~\ref{fig:cross1_h} and \ref{fig:cross2_h} show the cross-correlations of six combinations of different critical points. We will first describe the left panels of the figures and then discuss the right panels for cross-correlations on BAO scales.

We first show in Figure~\ref{fig:cross1_h} the cross-correlations between overdense and underdense critical points. The main feature of these four cross-correlations is a strong exclusion zone (see left panels), where the correlation function $\xi$ is close to $-1$. Note that our definition of the exclusion zone is rather strict \citep[for comparison, see][]{baldauf+16}. The corresponding exclusion zone radius, $r_\mathrm{ez}$, defined implicitly by $\xi(r_\mathrm{ez})= -0.99$ is displayed with dashed vertical line and given in Table~\ref{tab:exclu_pos_rarity} together with the inflection point position for different abundance. In contrast to the auto-correlations, at $r\ge r_\mathrm{ez}$, the correlation functions monotonically increase toward zero without displaying any maximum. The cross-correlations being always negative implies that overdense and underdense critical points are always anti-clustered at all scales.  No statistically preferred distance between these pairs of critical points can be established.

The presence of the exclusion zone is always expected for the cross-correlation of critical 
points which signatures differ by more than one, as well as not differ at all as in the case of
auto-correlations. From curvature continuity argument, only critical points with signature differing strictly by unity, namely ${\cal PF}$, ${\cal FW}$, ${\cal WV}$ can approach each other when smoothly deforming the field. 
For the rest, the curvature has to change a sign at least 
twice along the line between the critical points (for instance, for two peaks, the density has to
go down and then up between them) which results in statistical exclusion of near pairs.
However, in Figure~\ref{fig:cross1_h} another effect also leads to the repulsion between close pairs, which is 
our choice to limit the rarities of peak and filament points to the highest (positive) end, whereas voids and walls are constrained to the lowest (negative) side. This difference of allowed density values between the overdense and underdense critical points produces the exclusion zone which can be attributed to a density continuity argument (it is highly unlikely to have large variation of density over short distances) even 
in the case of ${\cal FW}$ where the curvature repulsion effect is absent. As  expected,  the ${\cal FW}$ cross-correlations show slightly smaller exclusion zone, compared to those with a larger difference of signatures (${\cal FV}$ and ${\cal PW}$, then ${\cal PV}$). 
This is already captured in Gaussian random fields as shown in Figure~\ref{fig:crossGRF}.

\begin{table}
	\centering
	\caption{Exclusion zone radius $r_{\rm ez}$ and position of inflection point $r_{\rm inf}$ in cross-correlations as a function of rarity for the lowest redshift bin, $0\le z<0.5$. The errors are the standard deviations of the mean obtained from the resampled realisations at $z=0$ using a bootstrap method. 
	}
	\label{tab:exclu_pos_rarity}	
    \begin{tabular}{cccc} % six columns, alignment for each
        \hline\hline
		type & abundance & $r_{\rm ez}$ [\mpcph] & $r_{\rm inf}$ [\mpcph]\\
		\hline
		$\mathcal{PW}$
		& $20\%$ & $18.7\pm0.03$ & $132.4\pm0.9$\\
		& $15\%$ & $19.4\pm0.05$ & $132.6\pm1.0$\\
		& $10\%$ & $20.3\pm0.07$ & $133.3\pm1.2$\\
		& $5\%$ & $21.6\pm0.11$ & $135.2\pm4.4$\\
		\hline
		
		$\mathcal{PV}$
		& $20\%$ & $20.1\pm0.05$ & $132.3\pm2.4$\\
		& $15\%$ & $20.7\pm0.05$ & $133.2\pm3.9$\\
		& $10\%$ & $21.4\pm0.07$ & $134.8\pm4.3$\\
		& $5\%$ & $22.6\pm0.19$ & $138.2\pm3.2$\\
		\hline
		
		$\mathcal{FW}$
		& $20\%$ & $17.2\pm0.02$ & $134.5\pm1.2$\\
		& $15\%$ & $18.1\pm0.04$ & $134.7\pm1.1$\\
		& $10\%$ & $19.1\pm0.04$ & $135.0\pm1.2$\\
		& $5\%$ & $20.5\pm0.09$ & $137.8\pm1.7$\\
		\hline

		$\mathcal{FV}$
		& $20\%$ & $18.6\pm0.03$ & $134.6\pm2.1$\\
		& $15\%$ & $19.2\pm0.05$ & $135.3\pm2.6$\\
		& $10\%$ & $20.1\pm0.06$ & $135.7\pm2.8$\\
		& $5\%$ & $21.2\pm0.13$ & $137.9\pm3.4$\\
		\hline
    \end{tabular}
\end{table}

The  size of the exclusion zone -- meaning the volume where the probability to have both rare overdense and rare underdense critical points is suppressed -- grows with rarity. This is expected since a lower abundance threshold implies that overdense and underdense critical points have an even larger difference of rarity ($\Delta\nu$) which forces them to be further apart (the correlated density field will need a larger typical distance to be able to go from one very underdense region to one very overdense).
This results in the increase of the exclusion zone as abundances go to lower percentage values. 
Appendix~\ref{sec:cross-rare} shows how the rarity constraints change the behaviour of the cross-correlations at small separation, in particular as rarity threshold varies.   

As shown in the left panel of Figure~\ref{fig:cross2_h}, the cross-correlations between critical points with the same over-density signs are fundamentally different from those in the left panels of Figure~\ref{fig:cross1_h}. Firstly, they diverge at zero separation and exhibit neither exclusion zones nor negative correlations at small scales. Secondly, these cross-correlations have a local maximum at $r\approx13-15$\mpcph, similar to the maximum in the auto-correlation functions in Figure~\ref{fig:auto_h}. \footnote{Note that we have truncated the peak-filament and the wall-void correlation functions below $r<5$\mpcph, since the diverging feature at that scale is not fully captured by our measurements given the resolution limit of the density field. It is noteworthy that a 
similar divergence exists at small separation in the theoretical Gaussian prediction, for which there is no numerical resolution issue, as can be seen in Figure~\ref{fig:pf-abundancesGRF}.}  The reason for this divergent behaviour
at $r \to 0$ of peak-filament and wall-void cross correlations is that
in these cases we compare two critical points which signatures differ only by one (such that there is no need to go through one or more intermediate critical point in between, as is the case when going from peak to wall through filament for instance) and which ranges of allowed density values overlap.
Such critical points can come together at $r\to 0$ in merging \textit{critical events} \citep{cadiou2020}.
Possibility to have critical events induces positive correlation and enhanced probability of a critical pair with neighbouring signatures to be within a  shrinking volume of radius $r \to 0$.
This argument would extend to filament-wall cross-correlation, however
this case is modified by our choice of rarity thresholds which leads to
non-overlapping density values for filaments and walls in all pairs, which precludes
the merging critical events. Thus, this case is not included in Figure~\ref{fig:cross2_h}, but appeared in  Figure~\ref{fig:cross1_h}
and is discussed in more details in Appendix~\ref{sec:cross-rare} (see Figure~\ref{fig:cross_FW+}). 

The enhanced probability of high filamentary saddles to exist near high peaks shown in peak-filament cross-correlations matches the topological properties of the Cosmic Web where
high peaks are the nodes for filamentary branches (that eventually pass through saddles). The presence of the local maximum in their cross-correlation function
points to some order in spatial distribution of saddles around the
peaks, and gives a statistically preferred distance to such saddles. The typical length of filamentary bridges between two high peaks can be estimated to twice this distance.
The position of the local maximum, interestingly, varies negligibly with abundance, at $r_{\rm max} \approx 14$\mpcph\, leading to an estimate
of 28\mpcph\ for a typical bridging filament length. 
It is remarkably close to the 30\mpcph\ estimate given in \cite{bkp96}.

The void-wall cross-correlation is very similar to the peak-filament one (and is exactly the same for Gaussian random fields).
Walls encompass the voids and define their boundary.
Let us recall that by rare walls we mean the least dense ones, so our
rare wall saddles are the points through which voids percolate first.
According to our measurements, the preferred distance to them designated by the position of 
the local correlation maximum is $r_{\rm max} \approx 15$\mpcph. It can potentially be interpreted as the typical largest radius of a sphere centred at the void extremum that can be inscribed into the void boundaries. 

We also note that the correlation amplitude shows a clear dependence on the critical point abundances, increasing with rarity. This points towards
an even stronger statistical regularity in the relative spatial distribution of the very rare peaks and filaments, and voids and walls.

At the mildly non-linear stage ($0\le z<0.5$) exhibited in Figure~\ref{fig:cross2_h}, we find a higher maximum for peak-filament cross-correlations than for the wall-void case, indicating that the clustering of the overdense pairs are stronger than that of their underdense counterpart with the same abundance. This is an effect of the non-linear evolution, since in
the linear Gaussian limit ${\cal P\!F}$ and ${\cal W\!V}$ pairs
are exactly symmetric. We shall look at this non-linear evolution in more detail
in the following section.

When compared to the Gaussian case in Figure~\ref{fig:crossGRF}, the measured cross-correlations in the left panels of Figures~\ref{fig:cross1_h} and \ref{fig:cross2_h} are qualitatively consistent with the predictions. Cross-correlations between peaks and underdense points (\ie walls or voids) are increasing functions of the radius, going from $\xi=-1$ to $\xi=0$. The exclusion zone is clearly visible until radius up to $\sim 3R_{\rm G}$ ($\sim 18$\mpcph). The size of the exclusion zone (slightly) increases similarly to the measurements as $r_{{\rm ez},{\mathcal PV}}>r_{{\rm  ez},{\mathcal PW}} \approx r_{{\rm  ez},{\mathcal FV}}>r_{{\rm ez},{\mathcal  FW}}$, where $r_{{\rm ez},{\mathcal PV}}$, $r_{{\rm ez},{\mathcal  PW}}$, $r_{{\rm ez},{\mathcal  FV}}$, and $r_{{\rm ez},{\mathcal  FW}}$ represent the radius of the exclusion zone for peak-void, peak-wall, filament-void, and filament-wall, respectively.
While the existence of the exclusion zone stems from the curvature and density differences, let us emphasize again that the ordering can be explained through a geometrical argument.
Indeed, one can go directly from a filament to a wall, while going from a peak to a wall (resp. from a filament to a void) requires going through a filament (resp. a wall).
We therefore expect the exclusion zone of filament with walls to be less extended than that of peaks with walls and filaments with voids.
With a similar argument, we expect the largest exclusion zone to be found for peak-void pairs, since going from one to the other requires going through a wall and a filament.

Let us now move to the right panels of Figures~\ref{fig:cross1_h} and ~\ref{fig:cross2_h} which focus on BAO scale. On these large scales, the acoustic features appear very different in the two figures although always broader than in the dark matter correlation function. For points with different density signs (Figure~\ref{fig:cross1_h}), the BAO signature is shown as a dip in all cross-correlations, whereas it corresponds to a bump for points with the same density sign (Figure~\ref{fig:cross2_h}) including the previous auto-correlations. The BAO dip in the right panels of Figure~\ref{fig:cross1_h} actually represents a stronger anti-clustering at the BAO scale since $\xi$ is negative until $r\approx133$\mpcph. This is a direct consequence of the fact that the  voids and walls we consider have a negative rarity $\nu$ whose corresponding bias factor is negative  \citep[see \eg figure 9 of][]{Uhlemann:2016un}. Interestingly, the inflection points occurs again at about $133$\mpcph\, which is consistent with the auto-correlations (see Table~\ref{tab:exclu_pos_rarity}).

As for the cross-correlations of critical points with the same density sign, the BAOs appear as a bump (as the bias factors have the same sign in this case hence their product is positive) whose position is poorly defined and noisy (right panels of Figure~\ref{fig:cross2_h}). 
As can be seen on these plots, we detect a stronger dependence on the abundance for the peak-filament correlations. This is consistent with the results for the auto-correlations that were showing relatively larger amplitude changes on very large scales in the peak-peak and filament-filament cases than the rest of the auto-correlations.
Similarly to the previous correlation functions, the location of the inflection point is robustly at $\sim133$\mpcph\ (see Table~\ref{tab:peak_posh_rarity}). Note interestingly that all kinds of auto- and cross- correlations have inflection point consistently at the same scale $r_{\rm inf}\simeq133$\mpcph.

\begin{figure*}
	\includegraphics[trim=0 120bp 0 120bp,clip,width=2\columnwidth]{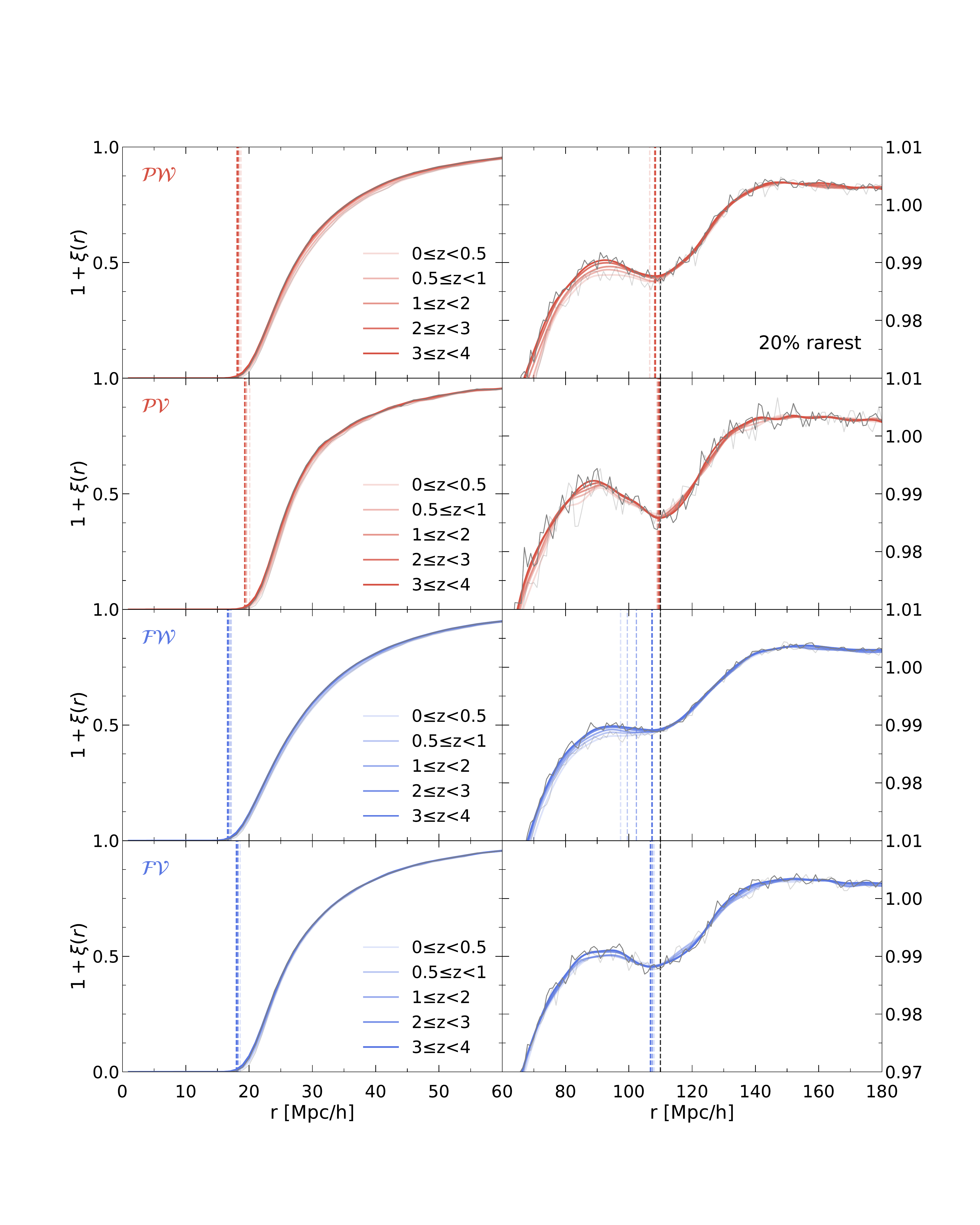}
    \caption{Following Figure~\ref{fig:cross1_h}, the redshift evolution of the 
    cross-correlations of the 20\% rarest critical points  for separations between 0 and 60\mpcph\ ({\it left}) and BAO scales ({\it right}). The ${\cal PW}$ (peak-wall), ${\cal PV}$ (peak-void), ${\cal FW}$ (filament-wall), and ${\cal FV}$ (filament-void) correlations are shown from top to bottom. The vertical lines mark the exclusion zone radius ({\it left}) and the positions of BAO ({\it right}) of the cross- (non-black) and the linear matter correlation (black). The size of the exclusion zone negligibly changes with time compared to Figure~\ref{fig:cross1_h}. Interestingly, little redshift evolution of the cross-correlations is seen at large separation, although auto-correlations involving overdense critical points show a noticeable redshift variation. In most cases, the dip position of the BAO are well defined and stationary, except in the $\cal F\!W$ case. Note that the BAO position in the linear dark matter case is very close to that in the cross-correlations.}
    \label{fig:cross1_z}
\end{figure*}

\begin{figure*}
	\includegraphics[trim=0 30bp 0 30bp,clip,width=2\columnwidth]{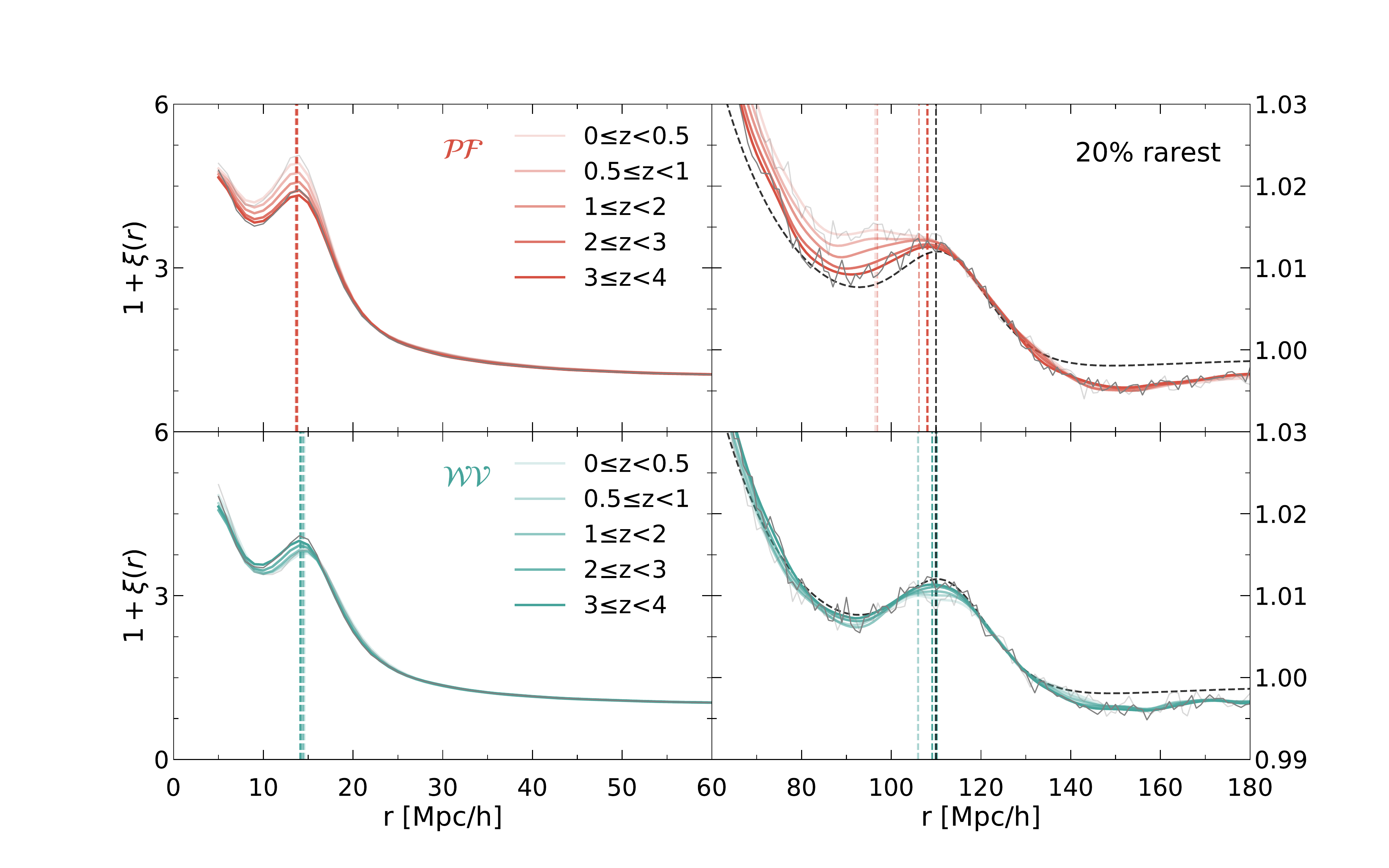}
    \caption{Same figure as Fig.~\ref{fig:cross1_z}, focusing on the redshift evolution of 
    the cross correlation of ${\cal P\!F}$ (peak-filament) and ${\cal W\!V}$ (wall-void)  in  the top and bottom panels, respectively. The linear dark matter correlation function (dashed) at the median redshift of $0\le z <0.5$ (multiplied by a factor of 8) is also presented. The vertical lines mark the positions of the maximum ({\it left}) and the BAO ({\it right}) of the cross-correlation (non-black) and the linear matter correlation (black). The maximum at $\approx 15$\mpcph\ barely shifts but its height changes with redshift. The height increases for ${\cal P\!F}$, whereas it decreases for ${\cal W\!V}$ with cosmic time. The BAO feature in the ${\cal P\!F}$ correlation becomes broader at lower redshift, whereas the BAO shape in the ${\cal W\!V}$ correlation negligibly changes. In general, the positions of the BAO bump for both ${\cal P\!F}$ and ${\cal W\!V}$ are stationary at all measured redshifts.
    }
    \label{fig:cross2_z}
\end{figure*}

\subsubsection{Redshift evolution of cross-correlations: Figures~\ref{fig:cross1_z}-\ref{fig:cross2_z}}\label{sec:cross-evolution}

Let us now turn to the redshift evolution of the cross-correlation functions. In Figure~\ref{fig:cross1_z}, we again first show peak-wall, peak-void, filament-wall, and filament-void cross correlations but for different redshift bins. We also provide the relevant quantities for the exclusion zone and the BAO features as functions of redshift in Table~\ref{tab:exclu_bao_redshift}. Similarly to the auto-correlations, the behaviour of the cross-correlations changes little with redshift. In the left panels, the correlation functions for different redshift bins almost overlap with each other at small separation. In particular, the exclusion zone size remains nearly constant with redshift 
such that they maintain the same ordering across cosmic time: $r_{\rm{ez},{\cal P\!V}}>r_{{\rm ez},{\cal P\!W}}\approx r_{{\rm ez},{\cal F\!V}}> r_{\rm ez,{\cal F\!W}}$.

\begin{table*}
	\centering
	\caption{Exclusion zone radii and positions and heights of BAO in cross- correlations as a function of redshift for $20\%$ abundance. The errors are the standard deviations of the mean obtained from the resampled realisations using a bootstrap method at the lowest redshift snapshot of each redshift bin.
	}
	\label{tab:exclu_bao_redshift}	
    \begin{tabular}{ccccc} % six columns, alignment for each
        \hline\hline
		type & redshift & $r_{\rm ez}$ [\mpcph] & $r_{\rm BAO}$ [\mpcph] & $h_{\rm BAO}$\\
		\hline
		$\mathcal{PW}$
		& $0\le z<0.5$ & $18.7\pm0.03$ & $106.6\pm3.8$ & $0.987\pm0.001$\\
		& $0.5\le z<1$ & $18.4\pm0.05$ & $108.1\pm1.4$ & $0.987\pm0.001$\\
		& $1\le z<2$ & $18.3\pm0.03$ & $108.2\pm1.0$ & $0.987\pm0.001$\\
		& $2\le z<3$ & $18.2\pm0.03$ & $108.3\pm1.3$ & $0.988\pm0.001$ \\
		& $3\le z<4$ & $18.1\pm0.05$ & $108.4\pm0.9$ & $0.988\pm0.001$\\
		\hline
		
		$\mathcal{PV}$
		& $0\le z<0.5$ & $20.1\pm0.05$ & $108.8\pm2.7$ & $0.987\pm0.001$\\
		& $0.5\le z<1$ & $19.6\pm0.05$ & $109.1\pm1.3$ & $0.986\pm0.001$\\
		& $1\le z<2$ & $19.4\pm0.05$ & $109.3\pm1.5$ & $0.986\pm0.001$\\
		& $2\le z<3$ & $19.3\pm0.05$ & $109.5\pm1.0$ & $0.986\pm0.001$ \\
		& $3\le z<4$ & $19.3\pm0.05$ & $109.8\pm0.8$ & $0.986\pm0.001$\\
		\hline
		
		$\mathcal{FW}$
		& $0\le z<0.5$ & $17.2\pm0.02$ & $97.4\pm3.3$ & $0.988\pm0.001$\\
		& $0.5\le z<1$ & $17.1\pm0.05$ & $99.5\pm3.8$ & $0.989\pm0.001$\\
		& $1\le z<2$ & $16.8\pm0.04$ & $102.4\pm3.0$ & $0.989\pm0.001$\\
		& $2\le z<3$ & $16.7\pm0.01$ & $107.4\pm2.7$ & $0.989\pm0.001$ \\
		& $3\le z<4$ & $16.6\pm0.05$ & $107.3\pm2.0$ & $0.989\pm0.001$\\
		\hline
		
		$\mathcal{FV}$
		& $0\le z<0.5$ & $18.6\pm0.03$ & $108.0\pm3.0$ & $0.989\pm0.001$ \\
		& $0.5\le z<1$ & $18.3\pm0.05$ & $107.1\pm1.4$ & $0.988\pm0.001$ \\
		& $1\le z<2$ & $18.2\pm0.05$ & $107.5\pm0.9$ & $0.988\pm0.001$ \\
		& $2\le z<3$ & $18.1\pm0.03$ & $106.9\pm1.5$ & $0.988\pm0.001$ \\
		& $3\le z<4$ & $18.0\pm0.04$ & $106.8\pm0.9$ & $0.988\pm0.001$\\
		\hline
    \end{tabular}
\end{table*}

In the right panels, the BAO feature is evident in all redshift ranges. The BAO positions are stable at all redshifts and are close to that of the (unsmoothed) linear matter correlation in most cases, except the filament-wall correlation for which the wiggle gets smeared and closer to a plateau-like shape probably because those critical points are the least biased and therefore cannot overcome the effect of the smoothing thanks a curvature effect. The height at the BAO position is also redshift independent, even for the filament-wall case. These four cross-correlations are better for analysing the BAO feature than the auto-correlations since they are well-defined and more stable throughout the redshift evolution.

Figure~\ref{fig:cross2_z} shows the redshift evolution of the peak-filament and wall-void correlations. As can be seen in the left panels, the region around the local maximum shows the strongest redshift evolution, in particular when compared to the diverging feature at small separation ($r<8$\mpcph). Notably, the {$\cal P\!F$} correlation develops a higher maximum
at late time, whereas the height of the maximum is monotonically decreasing with time for the {$\cal W\!V$} case (see Table~\ref{tab:peak_bao_redshift}).
Thus the non-linear evolution of overdense pairs and underdense pairs diverges from their initially symmetric Gaussian state and their redshift evolution goes in the opposite direction:
 the overdense pairs become more clustered while underdense ones become slightly more uniformly distributed.  However, note that the redshift variation in the mildly
non-linear regime probed here is 
weak in the  {$\cal W\!V$} cross-correlations and only somewhat more 
noticeable in the {$\cal P\!F$}.
While the height of the maximum changes a bit, its position changes even less so with redshift. The small and intermediate scale behaviour of the measured  {$\cal P\!F$} correlation qualitatively agrees with the Gaussian case  in Figure~\ref{fig:crossGRF}.
In particular, note that the positions of the maximum for the measurement and the Gaussian prediction are about $r\simeq2.3R_{\rm G}$ ($\approx 14$\mpcph) and $r\simeq2.2R_{\rm G}$, respectively, which shows that the Gaussian predictions for the maximum position remain relevant down to the mildly non-linear regime.

At large separation, a clearer signature of BAO is seen in the  {$\cal W\!V$} correlation compared to the {$\cal P\!F$} case. The {$\cal W\!V$} correlation function shows negligible redshift evolution compared to the peak-filament case. Given the smearing of the BAO feature, the identification of the acoustic scale at late time for {$\cal P\!F$} is relatively less straightforward, compared to the {$\cal W\!V$} correlation.

\begin{figure*}
	\includegraphics[trim=0 20bp 20bp 0,clip,width=2\columnwidth]{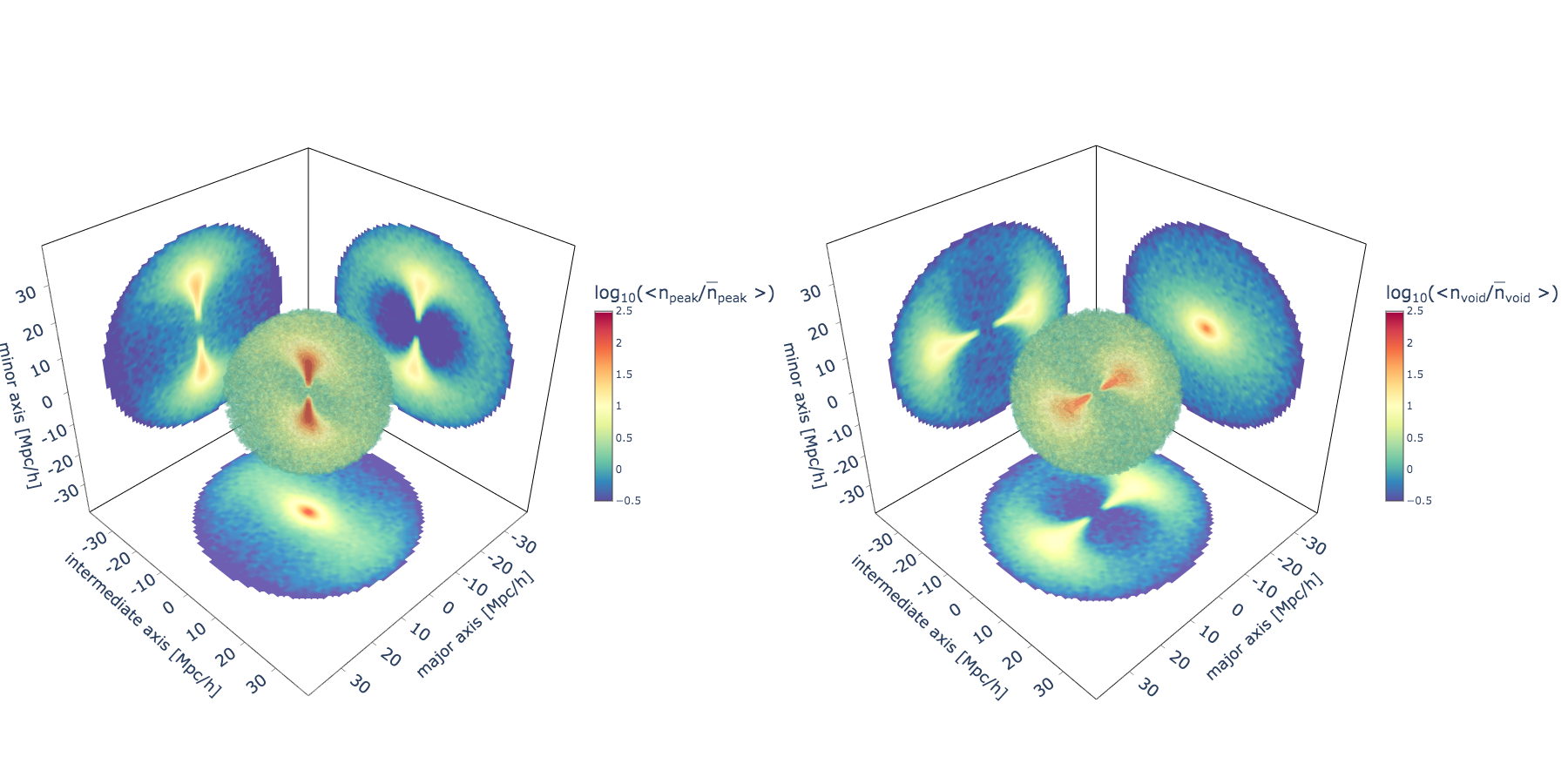}
    \caption{Overdensities of the rarest 20\% extrema in the eigen-frame of their (20\% rarest) saddle-points, together with their projection on planes perpendicular to the principal axes of the  Hessian, within $30$\mpcph\ of the saddle, as given by equation~(\ref{eq:overdensities}). Peaks around filaments (left) and voids around walls (right) are shown. 
    Excess number densities of peak and void points are found along the minor and major axis of the frame of saddles. Note that the peak (resp. void) number density along this preferential direction (in red) abruptly stops in the vicinity of the saddle-point, because it is close to the smoothing scale (6\mpcph) and resolution (2\mpcph) limit. See Figure~\ref{fig:crossGRF2D} for its GRF theoretical counterpart (which does not suffer from this limitation).
   }
    \label{fig:3dpeakcount}
\end{figure*}

\begin{figure*}
    \includegraphics[width=2\columnwidth]{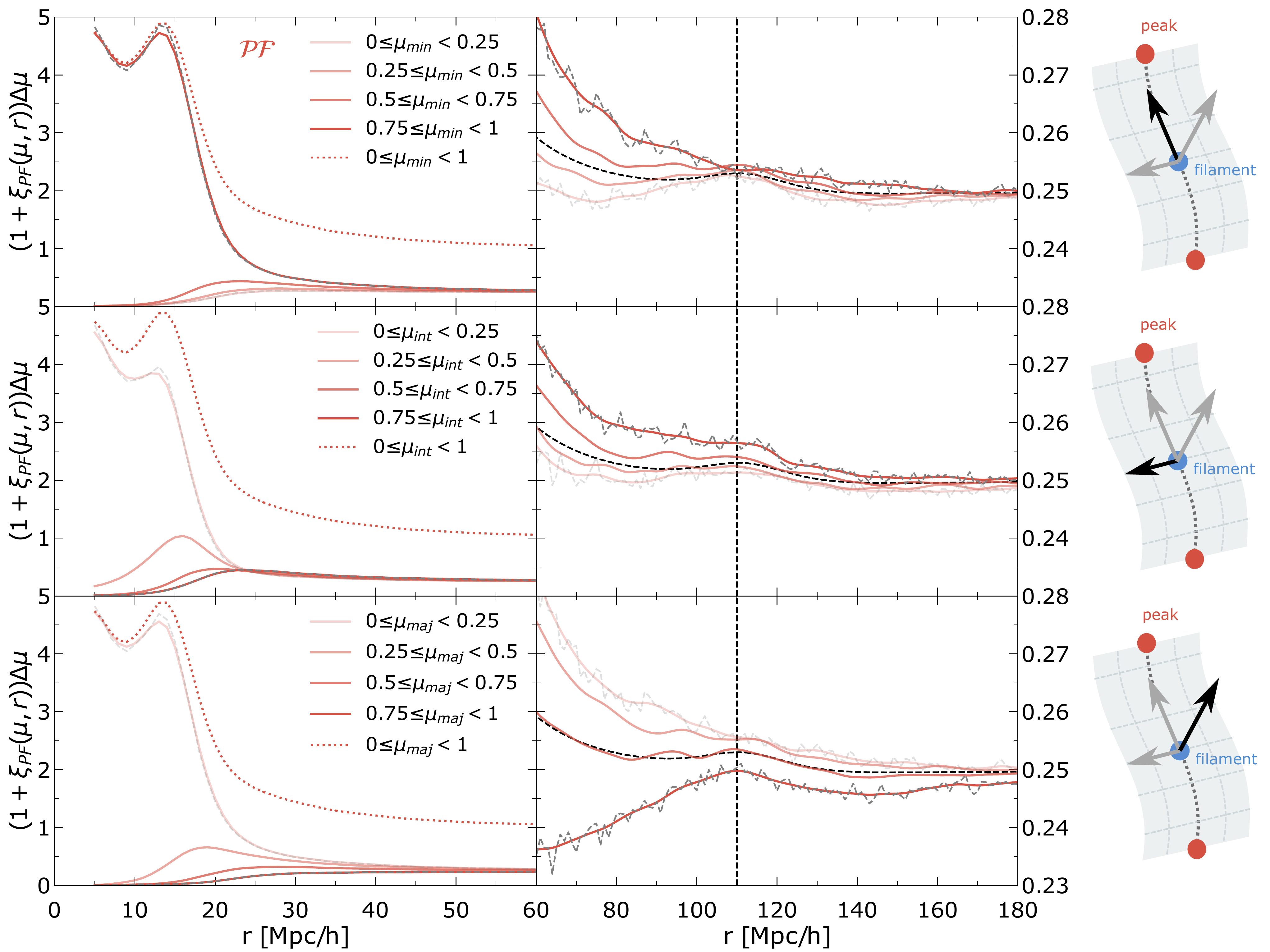}
    \caption{Directional impact on the $\cal P\!F$ cross-correlation function for the 20\% highest critical points at redshift $0\le z<0.5$ for separations between 0 and 60 Mpc$/h$ ({\it left}) and BAO scales ({\it right}), as given by equation~(\ref{eq:xi_dir}). The impact of the orientation of the separation vector with respect to resp. the minor, intermediate, and major axis of the filament's Hessian are presented in resp. the top, middle and bottom panels. 
    A schematic guides showing the reference's direction (black arrow)   and the other principal axes (grey arrows) at the saddle point (blue) defining the filament (dashed line connecting two peaks) are also depicted.
    We also display with a dashed solid line the linear matter correlation multiplied by the same factor of 8 as before and divided by 4 (which is the number of bins of angles). The vertical line marks the position of the BAO in the linear matter correlation. In the top panel, the small scale contribution to the ${\cal P\!F}$ correlation   comes mostly from the direction along the minor axis ($0.75\le \mu_{\mathrm{min}}<1$) of the eigen-frame, which implies that peaks are preferentially located along the filament's principal axis, as expected \citep{codis+18}. The rest of the orientations shows an exclusion-zone. For the middle (resp. bottom) panel, a major contribution at small separation originates from the direction perpendicular to the intermediate (resp. major) axis of the filament frame. This is qualitatively consistent with the 3D counts shown in Figure~\ref{fig:3dpeakcount}, left panel.
    At large separation, {it is still the direction along the minor axis of the filament that contributes most to the correlation in the top right panel. Although the correlation along the major axis is the minimum, it displays the most noticeable BAO feature at large separation.}
    }
    \label{fig:PFevec}
\end{figure*}

\begin{figure*}
	\includegraphics[width=2\columnwidth]{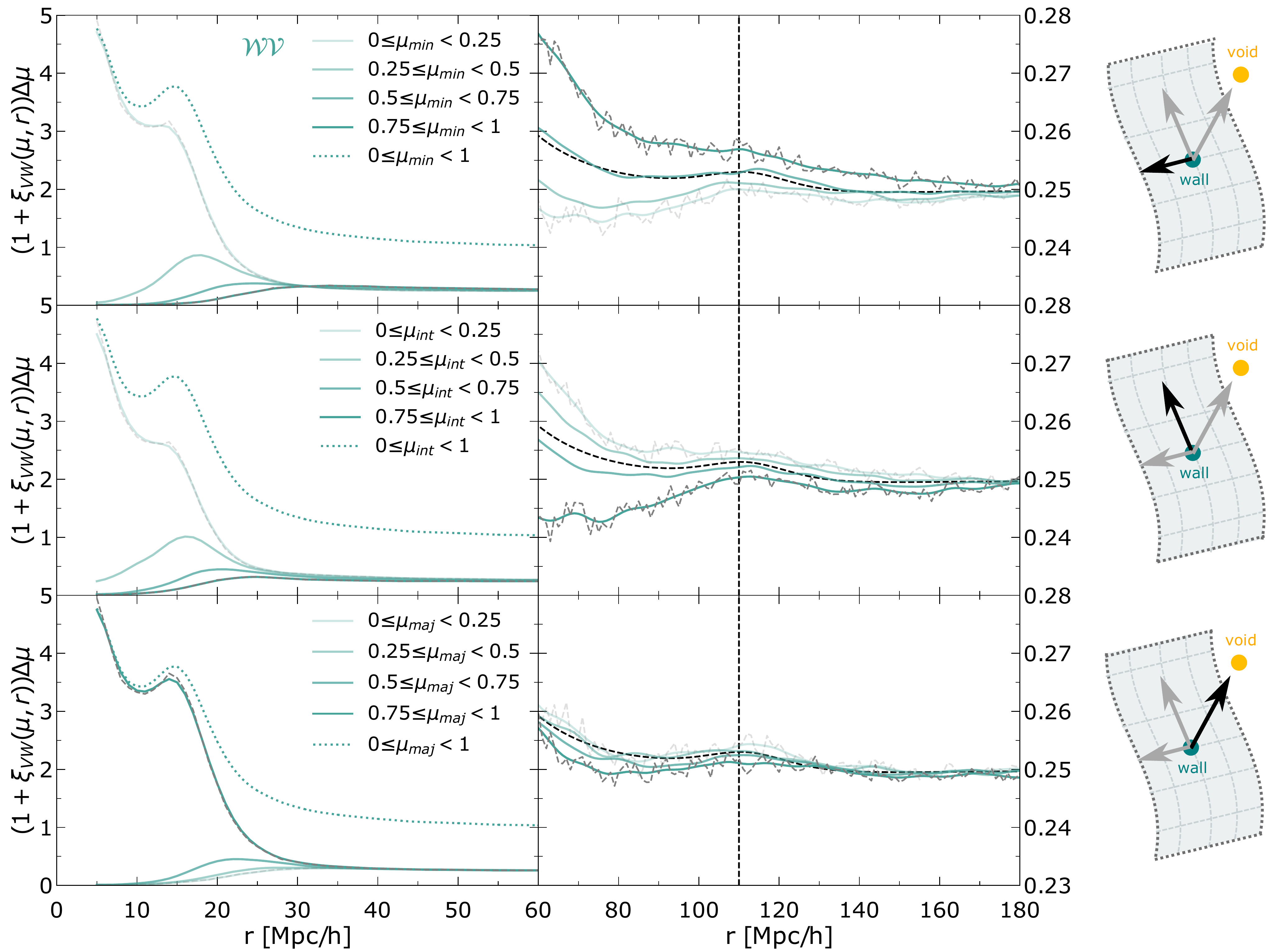}
    \caption{Directional contribution to the $\cal W\!V$ cross-correlation function of the 20\% lowest critical points for $0\le z<0.5$ for separations between 0 and 60 Mpc$/h$ ({\it left}) and BAO scales ({\it right}), as given by equation~(\ref{eq:xi_dir}). Different orientation with respect to the minor, intermediate, and major axis of the wall Hessian   are presented from the top to the bottom panel. A schematic guides showing the reference direction (black arrow) defining the  frame (grey arrows) are also depicted. We also display with a dashed solid line the linear matter correlation multiplied by the same "bias" factor of 8 as before and divided by 4 (which is the number of bins of cosines). The vertical line marks the position of the BAO in the linear matter correlation. In the bottom panel, the small scale contribution to the ${\cal W\!V}$ correlation mostly comes from the direction along the major axis ($0.75\le \mu{_\mathrm{max}}<1$) of the wall eigen-frame, which implies that voids are preferentially located around the wall-saddle point, in analogy to the  ${\cal P\!F}$ case (Figure~\ref{fig:PFevec}). See also the 3D counts  in Figure~\ref{fig:3dpeakcount}, right panel.
    At large separation, the BAO signature is more pronounced along the wall's intermediate axis.
 }    \label{fig:VWevec}
\end{figure*}

%%%%%%%%%%%%%%%
\subsubsection{Angle dependence of cross-correlations: Figures~\ref{fig:3dpeakcount}-\ref{fig:VWevec}} \label{sec:angle}
%%%%%%%%%%%%%%%%%%%%

To identify the origin of the characteristic features at $r<30$\mpcph\ in the peak-filament and wall-void cross-correlation functions, we investigate how the contribution to the correlation depends on the relative orientation of the pair in the eigenframe of the saddle defined by the principal axes of the local Hessian of the density field $H_{ij}=\partial_{ij} \delta$. 
Indeed, we expect a preferential alignment of the separation vector of the saddle-extrema pair in that frame, given that saddles bridge extrema points together along preferred directions \citep{Bond1996,Codis2017,codis+18}. In particular, we expect peak-filament pairs to cluster along the direction of positive curvature of the filament saddle and the void-wall pairs along the direction of negative curvature of the wall saddle. In what follows, 
we will denote $\lambda_1<\lambda_2<\lambda_3$ the three ordered eigenvalues of the Hessian matrix and $\vvec e_i$ their corresponding eigenvector. We will use the terminology minor axis for $\vvec e_3$, intermediate axis for $\vvec e_2$ and major axis for $\vvec e_1$ such that a filament joining a filament-type saddle and a peak is typically along the minor axis, $\vvec e_3$, corresponding to the eigendirection of positive curvature of the saddle.

To first get a global view, Figure~\ref{fig:3dpeakcount} plots the conditional peak (resp. void) overdensity in the 3D frame set by filaments (resp. walls)
\begin{equation}
\label{eq:overdensities}
\frac{\left\langle n_{\rm ext}|\mathrm{sad}\right\rangle}{{\overline n_{\rm ext}}}(\vvec r)\equiv
\frac{\left\langle n_{\rm sad}( \vvec r_{\rm sad}) n_{\rm ext}(\vvec r_{\rm sad}+{\cal R}\cdot \vvec r))\right\rangle_{\vvec r_{\rm sad}}}{{\overline n_{\rm ext}}{\overline n_{\rm sad}}},
\end{equation}
where $\vvec r$ is the separation vector in the eigenframe of the saddle,  $\vvec r_{\rm sad}$ is the position of a saddle point in the simulation frame on which we perform the average and $\overline n$ refers to the mean number density of points (extrema or saddles above threshold). The multiplication by the $3\times 3$ rotation matrix ${\cal R}=(\vvec e_1,\vvec e_2,\vvec e_3)$ rotates the position from saddle eigenframe to the Euclidean frame of the simulation.
For clarity, we also plot projections of this 3D function on the 2D planes perpendicular to each principal axis of a saddle.  

In the left panel, as expected, we find a larger overdensity of peaks along the minor axis, parallel to the eigenvector corresponding to the positive eigenvalue of the Hessian of the density field, $\lambda_3$, evaluated at the filament-type saddle point.
In particular, when peak overdensities are integrated along the minor axis, we clearly detect a high concentration of peaks along the axis of the filament, and a decreasing number density of peaks in the direction perpendicular to the filament.
This is in qualitative agreement with \eg \cite{codis2015,kraljic_galaxies_2018,Musso2018}
as well as \cite{bkp96} which took a peak-centred view arguing for enhanced conditional probability
of the filaments in the direction along the peak minor axis, especially between two aligned peaks.

The integration along the major axis confirms the existence of an exclusion zone with a dumbbell shape, showing a clear deficit in the projected peak number density on both sides of the filament and corresponding to two void regions, while this particular signature  is less significant when projected along the intermediate axis. This is because peaks are more likely to be found along the intermediate axis corresponding to a wall rather  than along  the major axis direction.
Conversely, we find an excess of voids along the major axis of the Hessian frame of wall-type critical points in the right panel. Similarly, the projected void number density on the plane spanned by the minor and intermediate axes of the wall frame shows a monotonic decrease in the void number density as a function of  distance from walls. On  other planes, we observe  clear exclusion zones, with dumbbell shapes centred on walls.

More quantitatively, Figure~\ref{fig:PFevec} (resp. Figure~\ref{fig:VWevec}) shows the anisotropic contribution to the cross-correlation from different directions identified with respect to each of the principal axis of the filament (resp. wall) Hessian's frame. 
We calculate the relative counts of critical point pairs with a particular orientation as
\begin{equation}
    (1+\xi_{ij}(\mu,r)) \Delta\mu = \frac{\langle C_{i}C_{j}(\mu,\mu+\Delta\mu)\rangle
    }{\sqrt{\langle C_{i}R_{j}\rangle \langle C_{j}R_{i}\rangle }}
    \sqrt{\frac{N_{R_j}N_{R_i}}{N_{C_j}N_{C_i}}},
    \label{eq:xi_dir}    
\end{equation}
where $\mu \in [0,1]$ is the cosine of the angle between the principal axis of the saddle frame and the separation vector of that pair, and $\langle C_{i}C_{j}(\mu,\mu+\Delta\mu)\rangle$ represents the number of saddle-extremum pairs whose directional cosine is in the range $[\mu,\mu+\Delta\mu]$.  
The sum of the contributions in equation~(\ref{eq:xi_dir}) for all angular bins is exactly the isotropic 2pCF of extrema and saddles.
The panels from top to bottom show binned number counts with angles starting from, resp., the minor, intermediate, and major axis of the saddle. The sum of all bins, identical for each panel, is also
shown. Let us remind that 
for filament saddles, the minor axis points in the most probable direction to the nearest peak,
while for wall saddles, the major axis points in the most probable direction of the nearest void (as represented in the schematic guides).

Let us first discuss the near behaviour of the anisotropic contributions to the cross-correlation\footnote{We should stress
that what is plotted are fractional contributions to the total correlation function from the pairs
within a given angular separation. As such individual curves should not be interpreted as
correlations themselves, in particular, the values below unity should not be automatically
interpreted as anti-correlation.}  for peaks and filaments, $ij=\mathcal{FP}$, presented in the left column of Figure~\ref{fig:PFevec}. 
Choosing the directional bin with the highest cosine values ($0.75\le \mu \le 1$) focuses on the close vicinity of a given saddle eigendirection. 
The bin with the lowest values ($0\le \mu < 0.25$) represents a disk, orthogonal to this
eigendirection.  

The structure of peak number distribution at distances
$r \lesssim 30\; h^{-1} \mathrm{Mpc}$ 
is seen to be strikingly different around different axes.
The small separation divergence and local maximum feature occur
specifically close to the minor axis direction.
In the directions misaligned with the minor axis,  peak-filament pairs  contribute significantly less than the aligned case, and even display some exclusion. 
This exclusion zone becomes most pronounced when one looks in the plane perpendicular to
the minor axis, either in the disk $0 \le \mu < 0.25$ in the top panel, or in the narrow cones
$0.75 \le \mu \le 1$ around intermediate and major axes in the middle and bottom panel.  This clear anisotropy suggests that peak critical points are distributed primarily along the minor axis  of the filament frame, as expected.
Along this direction, we expect to find peaks which can be arbitrary close to the filamentary saddle,
resulting in a divergence towards zero separation.  Figure~\ref{fig:PFevec} shows that this enhancement of peaks around filament comes
from a narrow angle range, $0.75\le \mu_{\rm min} <1$, around the minor axis.  
All other directions display some exclusion zone, more pronounced as $\mu$ increases.
A more careful analysis, performed on the Gaussian case, shows that the divergence as
$r \to 0$ of the $\mathcal{PF}$ cross-correlations exists only for a strictly tangential to minor axis approach to the saddle point $\mu=0$, all other directions displaying an anti-clustering region at small separation. When approaching the saddle point along a ray at a finite non-zero, but sufficiently small angle $\mu$, the $\mathcal{PF}$ correlation exhibits a local maximum  at $r \approx 20 h^{-1}\mathrm{Mpc}$ (corresponding to the crossing of the dumbbell contours described previously in the 3D case).
When integrated over angles, both features persist, leading to the characteristics shape for the full angle-integrated $\mathcal{PF}$ cross-correlation function discussed 
earlier which displays both a divergence at small separation and a local maximum.

The dependence of the peak-filament correlation on the orientation is in fact already fully captured in the initial Gaussian random field. Figures~\ref{fig:crossGRF2D} and~\ref{fig:crossGRF2D-nuth} show the  Gaussian versions of Figure~\ref{fig:PFevec}, together with the projection along the intermediate axis of the filament's frame as in Figure~\ref{fig:3dpeakcount}. We show the values of the Gaussian peak-filament correlations in a two-dimensional plane perpendicular to the intermediate axis of the filament frame. There is a good qualitative agreement between the measurement and the Gaussian prediction: stronger clustering along the minor axis, and an axi-symmetric exclusion zone near the minor axis. The angular contribution in the right panel of Figure~\ref{fig:crossGRF2D} also matches well the behaviour found in the measurement. The correlation decreases with increasing directional cosine defined with respect to the major axis of filament frame. In particular, the right panel further suggests that the divergence originates predominantly along the direction perpendicular to the major axis (presumably, along the minor axis) and the local maximum in the correlation results from the direction slightly less perpendicular to the major axis (presumably, slightly misaligned with the minor axis).

Note that the features on scales below $r\sim30$\mpcph\ also change significantly depending on the rarity constraints imposed on critical points. As shown in the left panel of Figure~\ref{fig:crossGRF2D-nuth}, for cross-correlations of peaks and filaments with different specific rarities, the exclusion zone becomes horizontally elongated, and disconnects high correlation region since the condition for the critical points to have different rarities makes it unlikely for them to be close to each other. In the right panel, the angular analysis again demonstrates the existence of an exclusion zone in all directions. The extent of the exclusion zone decreases with decreasing alignment to the major axis. On the other hand, the maximum amplitude increases with increased misalignment to the major axis, evidently showing an anisotropic distribution of peak critical points.

Let us now move to BAO scales. In the right panel of Figure~\ref{fig:PFevec}, the largest contribution to the cross-correlation still occurs along the minor axis of the saddle frame but also significantly along the intermediate axis. As the separation vector of a pair tilts away from the minor axis of the filament, the relative pair count decreases and falls below the isotropic linear-matter case, implying that peak and filament points are anti-clustered in all the directions perpendicular to the minor axis.
Interestingly, however, it is the major axis that shows the most noticeable BAO feature, as shown in the bottom right panel, which may support the idea that less dense directions may provide a more suitable environment for cosmological analysis.
We also note an inversion of the profiles along the intermediate axis (middle row): the highest relative counts at small separation becomes the lowest on BAO scales.

For the wall-void  counts presented in Figure~\ref{fig:VWevec}, $ij=\mathcal{WV}$, we find the opposite trend to the peak-filament counts: the main contribution to the divergence and the maximum is produced along the major axis of the wall frame, as can be seen in the bottom-left panel. We also detect an exclusion region in the relative counts, when the wall-void separation vector is misaligned ($0\le \mu_{\rm maj} <0.75$) with the major axis. Along the minor and intermediate axes, the pairs with the largest misalignment ($0\le \mu <0.25$) contribute the most. Void critical points are therefore typically located along the direction of the major axis of the wall frame, as expected, and creates a divergence at zero separation. The sum of this behaviour with exclusions along all the other directions creates the typical local maximum observed for the isotropic case. 
We once again note an inversion of the profiles, this time along the minor axis (top row): the highest relative counts at small separation becomes the lowest on BAO scales. We also note that for Gaussian random fields, this case is symmetric to the peak-filament correlation and therefore can be explained in a dual fashion. 

In the right panels of Figure~\ref{fig:VWevec}, the behaviour at large separation is different from the peak-filament case. Previously, the most significant BAO feature appeared along the major axis, although the dominant amplitude contribution to the isotropic correlation function came from the minor axis. In the present case, the BAO signature is more clearly detected along the intermediate axis, as can be seen in the middle right panel. Note that the largest contribution at small separation occurs along the major axis.

%%%%%%%%%%%%%%%%%%%%%%%%%%%%%%%%%%%%%%%%%%%%%%%%%%
\section{The average cosmic crystal}\label{sec:crystalMain}
%%%%%%%%%%%%%%%%%%%%%%%%%%%%%%%%%%%%%%%%%%%%%%%%%%
The features in the measured and predicted cross-correlation functions, and first of all
the existence of the clear local maximum of the cross-correlations at $r\approx13-15$\mpcph, indicate that critical points are not uniformly located but exhibit on average a regular pattern in their spatial distribution \citep{pichon2010,codis+18,cadiou2020}. 

Thus, we examine the relative position of critical points traced by the maxima of their cross-correlations to investigate how the critical points are spatially distributed by finding their characteristic clustering scale. Here we are interested
in the most prominent features of the Cosmic Web - densest peaks, filaments and walls.
We also select the densest voids as they may correspond to the interior of the densest walls and voids are often defined by their boundary. 
Here, in practice, we restrict the sample of critical points to their respective highest 20\% range.

\begin{figure}
	\includegraphics[trim=0 15bp 0 15bp,clip,width=1.1\columnwidth]{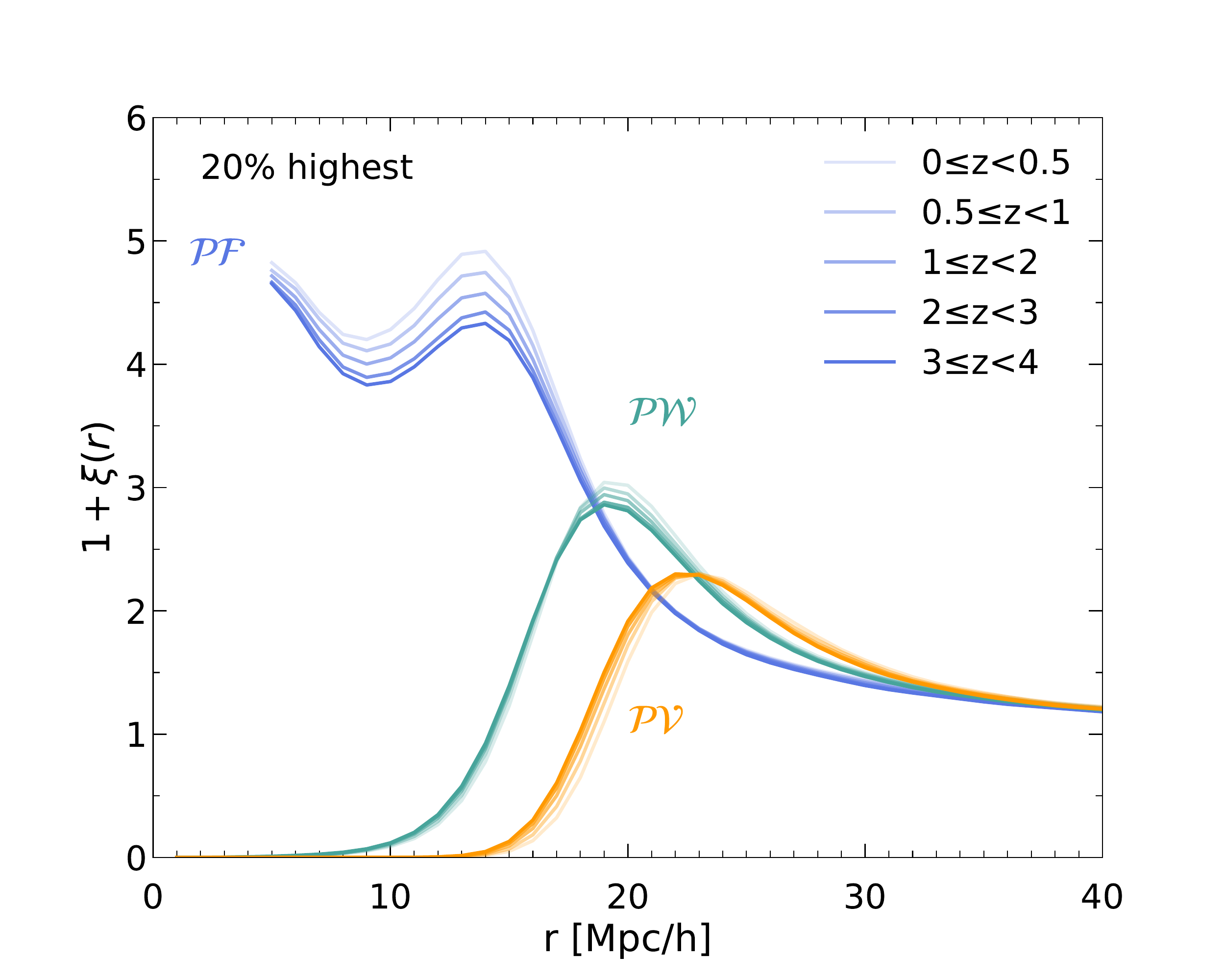}
	\caption{Redshift evolution of the correlations between peaks and different critical points as labelled, all in their 20\% highest rarity bin. Measured radii ratios are consistent with the Gaussian case shown in Figure~\ref{fig:crystal} as $r_{\cal P\!F}/r_{\cal P\!W}=0.7$ and $r_{\cal P\!F}/r_{\cal P\!V}=0.6$. 
	}
    \label{fig:crystal_measure}
\end{figure}

Figure~\ref{fig:crystal_measure} shows the redshift evolution of the correlations between peaks and the other critical points. 
We find that the typical peak-to-filament distance (as given by the position of the local maximum) is $\sim 0.7$ times the peak-to-wall most likely distance and $\sim 0.6$ times the peak-to-void separation. Remarkably, this is very close to the factors $1/\sqrt 2$ and $1/\sqrt 3$ expected in a cubic centred lattice.
These ratios remain nearly constant during the evolution. For the 5\% highest critical points, the ratios slightly increase to $r_{\cal P\!F}/r_{\cal P\!W}\approx 0.73$ and $r_{\cal P\!F}/r_{\cal P\!V}\approx 0.63$, but also in this case there is little redshift variation.
So it appears that the cosmic web through its evolution \textit{on average} bears similarities with a cubic lattice with the peak at the centre of the cube, filament-type saddles at the centre of the faces, wall-type saddles at the centre of the ridges and voids at the corners. 
This is consistent with the connectivity of the cosmic web in the initial conditions presented in Appendix~\ref{fig:crystal} and
investigated by \cite{codis+18}  \citep[see also][for observational evidence]{Darragh2019,kraljic_ImpactConnectivityCosmic_2020}. Note importantly that in practice the vicinity of each peak does not typically have this geometry: the regularity only arises because we investigate the mean field around them and therefore average out all fluctuations around it.  Most peaks are in fact dominated by one massive wall and two or three massive embedded filaments \citep{pichonetal11,codis+18}.

From a cosmological perspective, the relative insensitivity to redshift  evolution of   the ``cosmic crystal'' suggests it could possibly be used  as a  standard ruler, \eg to constrain the equation of state of dark energy, 
or to carry out an Alcock-Paczynski test \citep{2005MNRAS.364..743N,2011MNRAS.418.1725B,2019ApJ...881..146P}. 
Note that the crystal centred on the voids is dual to that centred on the peaks (see Figure~\ref{fig:crystal}, bottom panel).
Note finally  that  given the relative in-sensitivity to non-linearities of saddle points specifically,
it might be worth considering the clustering of saddles (Figures~\ref{fig:cross1_z} and~\ref{fig:cross_FW+})  as a more robust quantity to model. 
However, let us emphasise that the choice of considering the densest critical points (especially for voids) may not be the most realistic ones. We defer a thorough analysis of the cosmic crystal in realistic set-ups to future investigations. 
\footnote{In particular, an analysis of the crystal  as a function of the persistence of the cosmic web \protect{\citep{Sousbie2011}} would be an interesting avenue of research. }

%%%%%%%%%%%%%%%%%%%%%%%%%%%%%%%%%%%%%%%%%%%%%%%%%%
\section{Conclusions  and Perspectives}\label{sec:conclusions}
%%%%%%%%%%%%%%%%%%%%%%%%%%%%%%%%%%%%%%%%%%%%%%%%%%

\subsection{Conclusions}
%%%%%%%%%%%%%%%%%%%%%%%%%%
Focusing on both small-to-intermediate separations and BAO scales, the cosmic evolution of the 
clustering properties of peak, void, wall and filament critical points of fixed abundance was measured using two-point correlation functions  in $\Lambda$CDM dark matter simulations as a function of redshift. The two-point functions involving less non-linear critical points -- wall and filament saddle points -- proved more 
insensitive to redshift evolution, which may prove advantageous from a modelling perspective in the context of dark energy experiments. This conclusion is perfectly in line with the findings of
\cite{Gay2012} for one point statistics.
 
 A qualitative comparison to the corresponding theory for Gaussian Random fields (see Appendix~\ref{sec:GRF})  allowed us to understand the following features:
\begin{itemize}
\item The appearance of an exclusion zone  at small separation whose size increases with rarity and which depends on the signature difference between the critical points.
\item The divergence-and-maximum of cross-correlations of peaks and filament-type saddles (resp. voids and wall-type saddles), reflecting the relative  loci of such points along filaments (resp. walls), as expected from critical event theory \citep{cadiou2020}.
\item The amplification of the BAO bump with the rarity of the critical points for auto-correlations, which corresponds to a dip for cross correlations involving one negative bias factor.
\item The robustness of the observed location of the inflection point at $r\simeq 134\pm3$\mpcph\ corresponding to zero correlation in all pairs of critical points, which interestingly does not seem to depend on bias or redshift. 
\item The relative insensitivity to redshift  evolution of   the ``cosmic crystal'',  which suggests it could possibly be used  as a  standard ruler. 
\end{itemize}

The qualitative agreement between the simulations on the one hand, and the predictions  for Gaussian random fields on the other hand is striking  given the level of non-linearity probed ($\sigma\sim 0.6$).  This suggests that on the scales probed in this paper, the relative clustering of critical points was already encoded in the initial conditions \citep{Pogosyanetal1998} and is conserved throughout the cosmic evolution. 

Note that tables~\ref{tab:peak_posh_rarity} and \ref{tab:exclu_pos_rarity} quantified the characteristic scales and amplitude of these correlations as a function of rarity, and their redshift evolutions including BAO are summarized in Tables~\ref{tab:peak_bao_redshift} and \ref{tab:exclu_bao_redshift}, when relevant.

\subsection{Perspectives}
%%%%%%%%%%%%%%%%%%%%%%%%%%

As a first systematic investigation of  two points functions of other critical points (beyond peaks and void), this work is obviously non exhaustive. 
In this paper we only touched upon the properties of the two-point functions in the vicinity of the BAO scale,  in contrast to the vast literature  available for peaks, and the more limited work on voids \citep[\eg][]{2016PhRvL.116q1301K}. Extending some of this work to other critical points should be a priority in the future, in particular given 
their relative insensitivity to cosmic evolution.
Beyond the scope of this paper, it would be interesting to also quantify the two-point functions of critical points in
2D intensity maps  \citep{Madau1997,2015ApJ...803...21B,2017A&A...597A.136S} or weak lensing maps \citep{2018PhRvD..98d3526A}. It would also be of interest to extend the GRF predictions of Appendix~\ref{sec:GRF}
taking into account the Zeldovich displacements of critical points \citep{Regos1995} and/or introduce non-Gaussian corrections
\emph{via} a Gram-Charlier expansion of the joint PDFs, extending to two-point functions the results of \citet{Gay2012} and equation~\eqref{eq:edge}.
This would allow us to model the redshift evolution (Sections~\ref{sec:evolution-auto} and~\ref{sec:cross-evolution}) 
and construct the corresponding estimators for dark energy experiments. 
The density fields could then for instance be extracted from upcoming Ly-$\alpha$ tomographic  reconstructions \citep{pichon2001,2014ApJ...795L..12L} applied to wide field surveys like  PSF or WEAVE. 
It would eventually be worthwhile to also quantify the three-point  functions of critical points.
One could also investigate how modified gravity or primordial non-gaussianities impact the clustering of these critical points
\citep{2018MNRAS.475.3262F, 2019BAAS...51c..64D}.

\subsection*{Acknowledgements}
%%%%%%%%%%%%%%%%%%%%%%%%%%%%%%%%%%%%%%%%%%%%%%%%%%
JS is supported by a KIAS Individual Grant (PG071202) at the  Korea Institute for Advanced Study.
DP acknowledges a visiting fellowship from KIAS during his visit in Seoul when this project was initiated.
SC acknowledges support from the SPHERES grant ANR-18-CE31-0009 of the French {\sl Agence Nationale de la Recherche} and from Fondation MERAC.
CC is sponsored by the European Union Horizon 2020 research and innovation program, under grant agreement No. 818085 GMGalaxies.
This work has made use of the Horizon Cluster hosted by Institut d'Astrophysique de Paris. We thank Stéphane Rouberol for running smoothly this cluster for us.
This work was partially supported by the Segal grant ANR-19-CE31-0017 (\href{https://www.secular-evolution.org}{secular-evolution.org}) of the French {\sl Agence Nationale de la Recherche}.

\section*{Data availability}
The data underlying this article are available in the article and in its online supplementary material.

%%%%%%%%%%%%%%%%%%%% REFERENCES %%%%%%%%%%%%%%%%%%
%\bibliographystyle{mnras}
%\bibliography{reference} % if your bibtex file is called example.bib

%%%%%%%%%%%%%%%%% APPENDICES %%%%%%%%%%%%%%%%%%%%%

\appendix

%%%%%%%%%%%%%%%%%%%%%%%%%%%%%%%%%%%%%%%%%%%%%%%%%%
\section{Theory for GRFs}\label{sec:theory}
%%%%%%%%%%%%%%%%%%%%%%%%%%%%%%%%%%%%%%%%%%%%%%%%%%
\begin{figure}
	\includegraphics[width=1.\columnwidth]{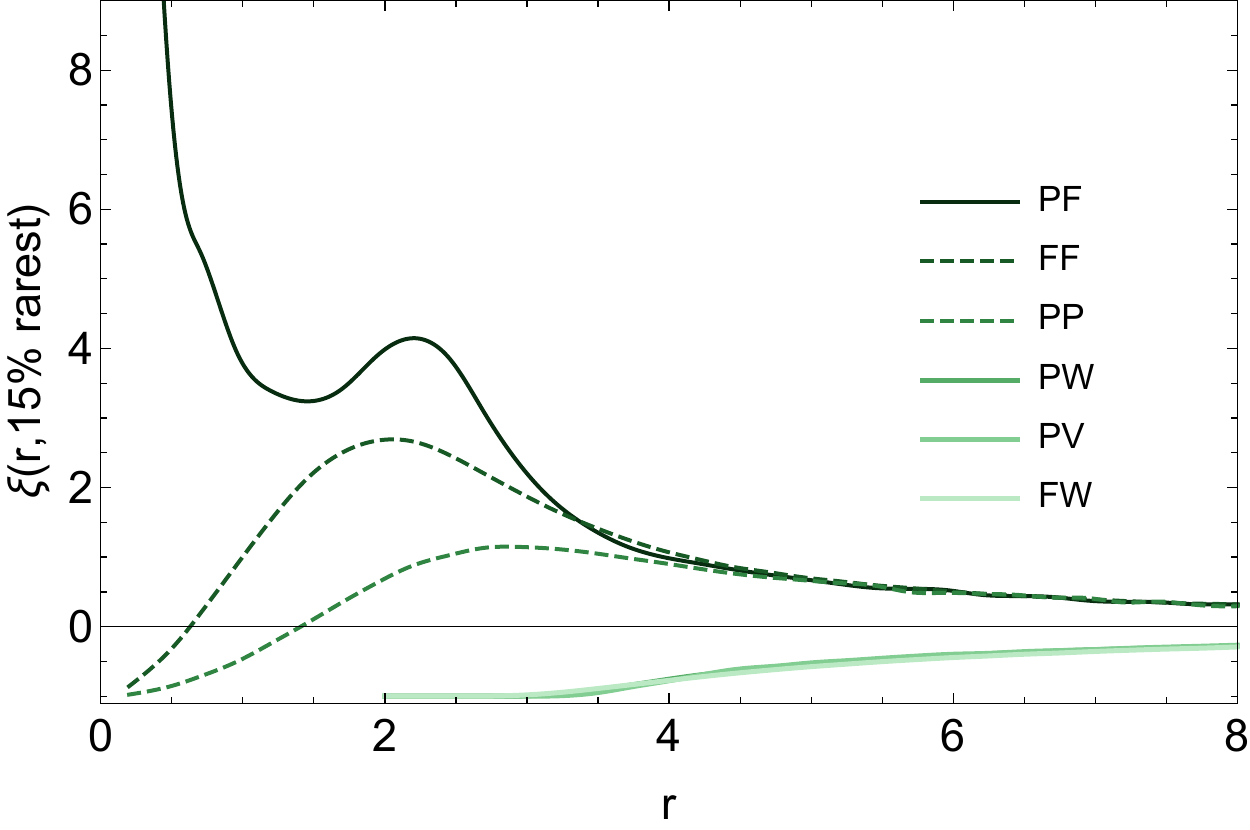}
    \caption{Auto (dashed) and cross (solid) 2pCFs as a function of the separation in units of the (arbitrary) smoothing length,  for GRF with $n_s=-1.5$ and for an abundance of $15\%$. Note that the curves for $\cal PW$, $\cal PV$ and $\cal FW$ correlations are almost indistinguishable on this plot.
 The correlations are qualitatively consistent with the measurements in simulations presented in the main text (Figure~\ref{fig:auto_h},~\ref{fig:cross1_h} and \ref{fig:cross2_h}).
}
    \label{fig:crossGRF}
\end{figure}

In order to enlighten the measurements presented in the main text,
let us  make Lagrangian predictions for the correlation of critical points in Gaussian random fields.
We will first predict the  auto- and cross-correlations for Gaussian random fields, and 
investigate the variation of the cross-correlation between peaks and filament-saddle points
in the frame of the hessian of the saddle as a function of orientation. Finally we will highlight 
the geometry of the so-called cosmic crystal which we can infer from these correlations.

\begin{figure*}
	\includegraphics[width=1\columnwidth]{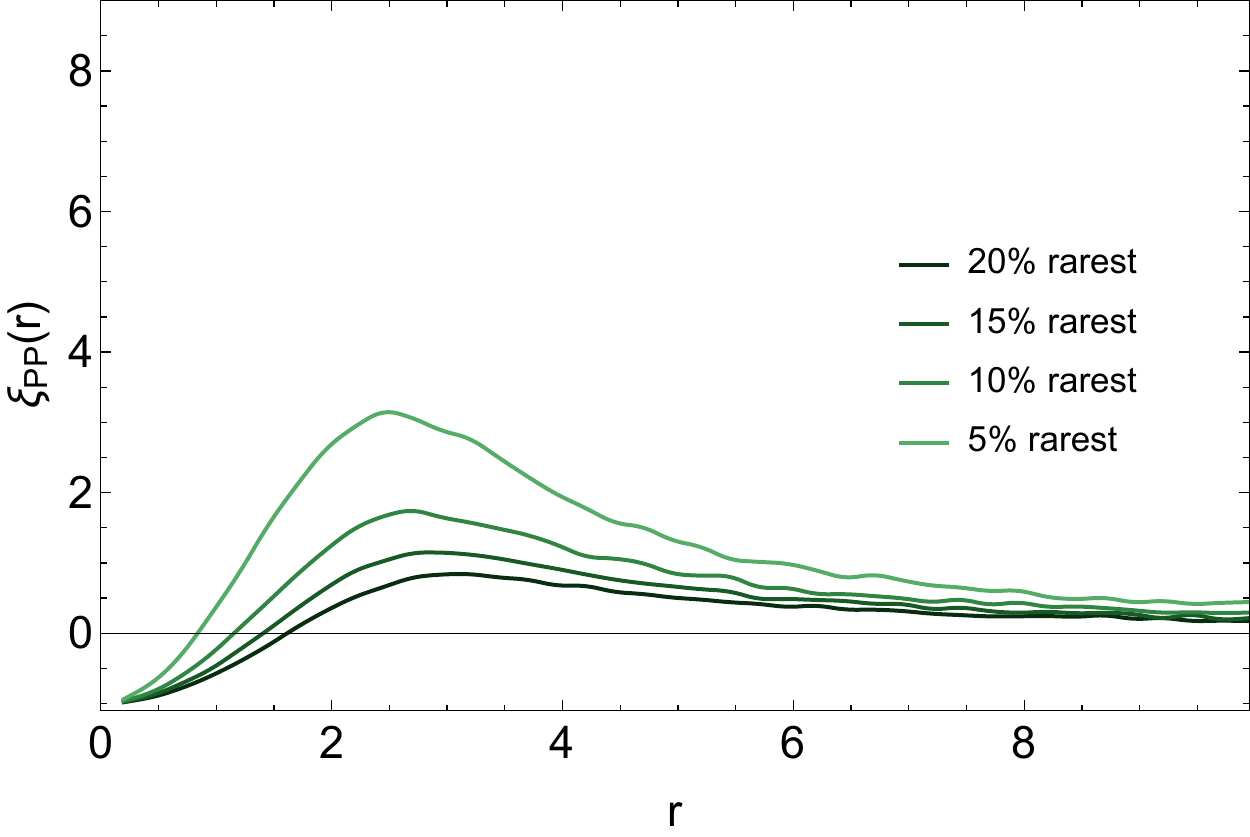}
\includegraphics[width=1\columnwidth]{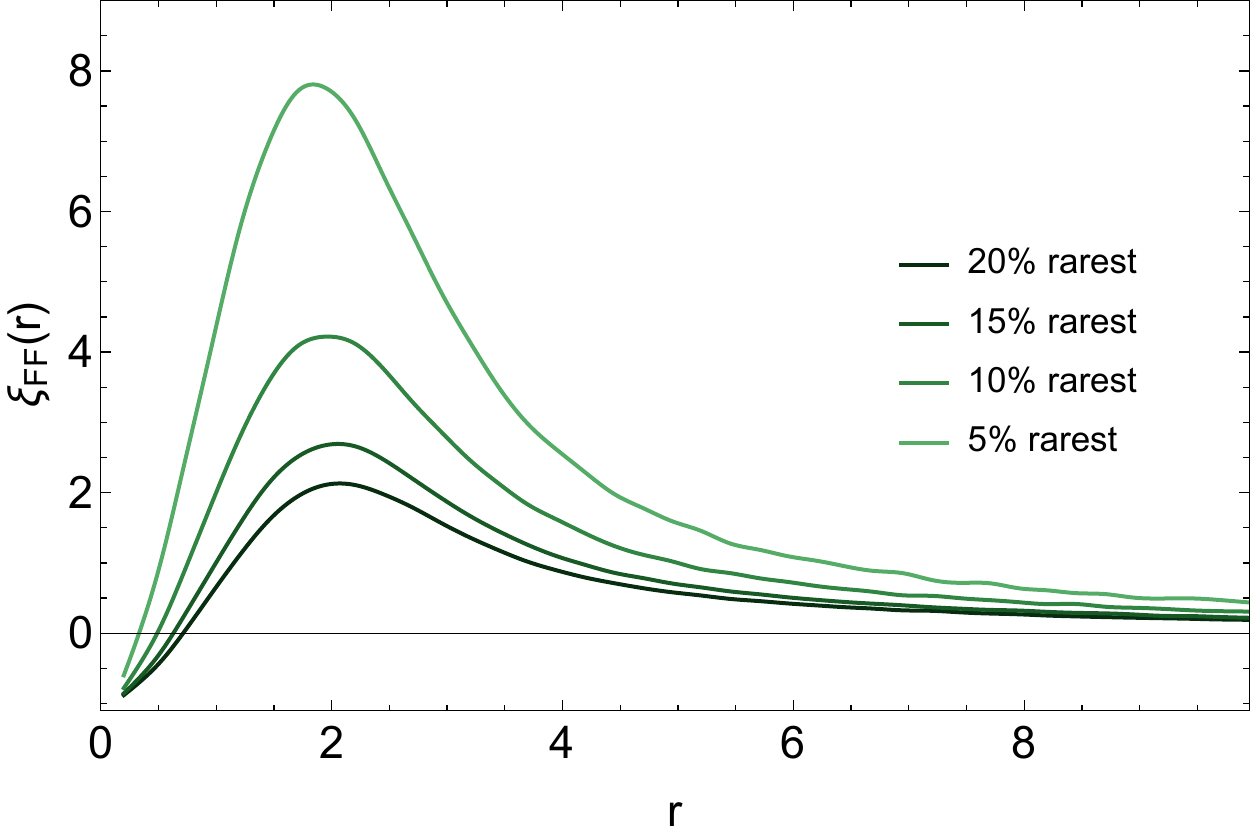}
    \caption{Auto 2pCFs (peaks on the left-hand panel, filaments on the right) for GRF with $n_s=-1.5$ and various abundances as labelled. The separation is in units of the smoothing length.  The  shape and amplitude of these functions  match  qualitatively the non-linearly evolved correlations presented in the main text (Figure~\ref{fig:auto_h}, top and middle left panels). As expected the rarer the critical points, the larger the amplitude.}
    \label{fig:autoGRF}
\end{figure*}

%%%%%%%%%%%%%%%%%
%\subsection{Clustering in GRFs}
\label{sec:GRF}
\subsection{Gaussian peak statistics}
Let us define the two-point correlation of all critical points based on the  joint statistics of the field  and its successive derivatives.
From the joint two-point statistics, we can define  the excess probability $\xi_{pq}(r,\nu_1,\nu_2)$  of having critical points of kind $p,q \in ({\cal P,F,W,V})$ of the field at height higher (for critical points with negative signature --peaks and filament--, lower for the ones with positive signatures --walls and voids--) than $\nu_1$ and $\nu_2$,  smoothed at scales $R$ and located at positions $(\rr_x, \rr_y)$,  as
\begin{align}
1+\xi_{pq}({r}|\nu_1,\nu_2) =\frac{\langle  {\rm cond}_p(\vvec{x})\times {\rm cond}_q(\vvec{y}) \rangle}{\langle  {\rm cond}_p(\vvec{x}) \rangle\langle  {\rm cond}_q(\vvec{x}) \rangle}\,,
\label{eq:jointcount}
\end{align}
where $\vvec{x}=\{x,x_i,x_{ij}\}$ (resp. $\vvec{y}$) is the set of fields at location $\rr_x$ (resp $\rr_y$), and
$r=|\rr_x-\rr_y|$ the spatial separation between the two points.

Here the field and its successive derivatives have been divided by their respective rms: $x= \delta/ \sigma_0$, $x_i=\nabla_i \delta/ \sigma_1$ and $x_{ij}=\nabla_i\nabla_j \delta/ \sigma_2$, where 
\begin{equation}
\sigma_i^2(R)\equiv \frac{1}{2\pi^2}\int_0^\infty \dd{k} k^2 P_k(k) k^{2i} W^2(k R)\,,
\label{eq:defsigi}
\end{equation}
(with $P_k$ the underlying power spectrum, and $W$ the Gaussian filter),
so that we have 
$\langle x^2\rangle = \sum_k\langle x_k^2\rangle =\langle ( \sum_{k} x_{kk})^2\rangle 
%= \sum_{k,l,m}\langle x_{klm}x_{klm} \rangle 
= 1$.

Evaluating the expectation in equation~\eqref{eq:jointcount} requires the full knowledge of the joint statistics of the field  ${ P}(\vvec{x},\vvec{y})$ (involving $(1+3+6)\times 2=20$ variables), given by
\begin{equation}
{P}(\vvec{x},\vvec{y})= \frac{\exp
\left[{\displaystyle-\frac{1}{2}}
\left(\begin{array}{c}
 \vvec{x}
\\
\vvec{y}
 \\
 \end{array} \right)^{\rm T}
 \cdot
  \mathbf{C}^{-1}\cdot \left(\begin{array}{c}
 \vvec{x}
\\
\vvec{y}
  \\
\end{array} \right) \right]
}
{{\rm det}|\left(2\pi\right)\mathbf{C}|^{1/2} } \,,
\label{eq:defPDF}
\end{equation}
where $\mathbf{C}$ is the covariance matrix which depends on the separation vector only because of homogeneity
\begin{equation}
\mathbf{C}=\left(\begin{array}{cc}
\mathbf{C}_{\mathbf{ xx}}&\mathbf{C}_{\mathbf{ xy}}\\
 \mathbf{C}_{\mathbf{xy}}^{\rm T}  &\mathbf{C}_{\mathbf{yy}}\\
\end{array}
\right)\,.
\end{equation}
Note that the correlation length of the various components of $\mathbf{C}_{\mathbf {xy}} $ differ, as higher derivatives decorrelate faster.
Equation~\eqref{eq:jointcount} evaluates the function $  {\rm cond}_i(\vvec{x})$, which for instance, for peaks, $p=P$, obeys \citep{BBKS} 
\begin{align}
 {\rm cond}_{P}(\vvec{x}) = & \dirac^{(3)}(x_i) |\mathrm{det}( x_{ik})|\heaviside(\!-\!\mathrm{det}( x_{ik}) )  \notag\\
 & 
\times \heaviside(\!-\!{\rm tr}( x_{ik}) ) \heaviside( {\rm tr}^2(x_{ik})  \!-\!{\rm tr} ( x_{il} x_{lk}))
\label{eq:cond2pt}
 \,, 
\end{align}
where the three Heaviside conditions ensure that  the determinant and the trace are negative while the minor
is positive so that the three eigenvalues are negative.  Similar conditions hold for the other critical points.

In practice we rely on Monte-Carlo
methods in {\sc Mathematica} in order to evaluate numerically equation~\eqref{eq:jointcount}. Namely, we draw random
numbers from the conditional probability  that $\vvec{x}$ and $\vvec{y}$ satisfy
the joint PDF~\eqref{eq:defPDF}, subject to the condition that $x_j=y_j=0$, 
$x>\nu_1$ and $y>\nu_2$ (resp. $<$ for walls and voids).
For each draw $(\vvec{x}^{(\alpha)},\vvec{y}^{(\alpha)})$, $\alpha=1,\dots,N$, we drop or keep the sample, depending on the signs of $x^{(\alpha)}-\nu$, $y^{(\alpha)}-\nu$ and the eigenvalues of $ (x^{(\alpha)}_{ik})$ and $ (y^{(\alpha)}_{ik})$.
If it is kept, we evaluate $\mathrm{det}\left( x^{(\alpha)}_{ik}\right) {\mathrm{det}}\left( y^{(\alpha)}_{ik}\right)$. Then, an estimate for the numerator of equation~\eqref{eq:jointcount} reads
\begin{equation}
\hskip -0.2cm\langle  {\rm cond}_p(\vvec{x}){\rm cond}_q(\vvec{y}) \rangle
\! \approx\! \frac{ \!{ P}_m(x_{i}\!=\!y_{k}\!=\!0)\!}{N}
\!\!\sum_{\alpha\in {\cal S}_{pq}} \!\! | \mathrm{det}( x_{ik}^{(\alpha)})\,\mathrm{det}( y_{ik}^{(\alpha)})|,
\nonumber
\end{equation}
where $N$ is the total number of draws, $P_m$ the marginal probability for the field  gradients, and ${\cal S}_{pq}$ is the subset of the indices of draws satisfying the constraints $p,q$ on the Hessians and the densities at separation $\rr$.
The same procedure can be applied to evaluate expectation entering  the denominator of  equation~\eqref{eq:jointcount},
\begin{equation}
\langle  {\rm cond}_p(\vvec{x}) \rangle
\! \approx\! \frac{ \!{ P}_m(x_{i}\!=\!0|r)\!}{N}
\!\sum_{\alpha\in {\cal S}_{p}} \! | \mathrm{det}( x_{ik}^{(\alpha)})|\,,
\nonumber
\end{equation}
which then yields an estimation of  $\xi_{pq}(r,\nu_1,\nu_2)$. This algorithm is embarrassingly parallel.

\begin{figure}
	\includegraphics[width=1.\columnwidth]{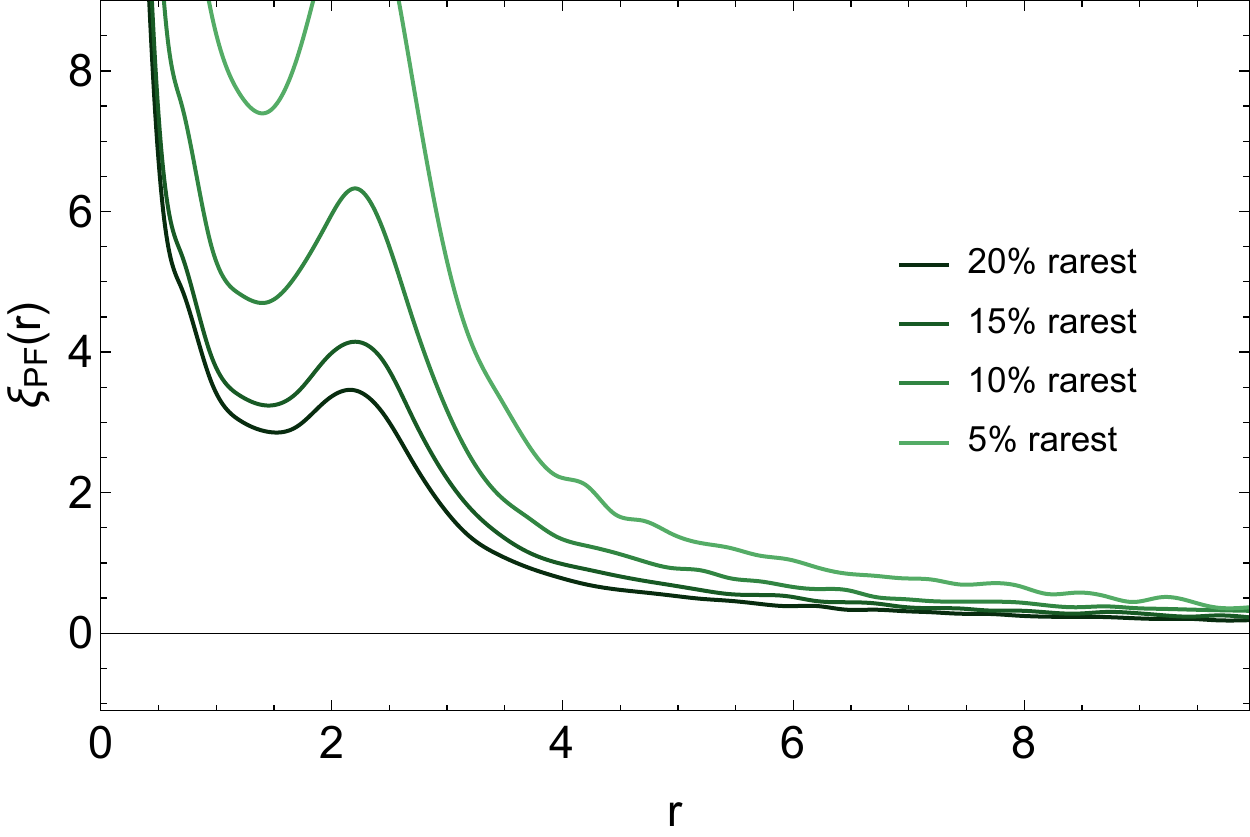}
    \caption{Peak-filament cross-correlations for GRF with $n_s=-1.5$ and various abundances as labelled. The separation is in units of the smoothing length.
    The  shape and amplitude of these functions also match  qualitatively their counterparts presented on (Figure~\ref{fig:cross2_h}, top left panel). }
    \label{fig:pf-abundancesGRF}
\end{figure}

\subsection{Isotropic clustering of critical points}
Figure~\ref{fig:crossGRF} shows predictions for all the cross-correlations and for an abundance of 15\% while Figure~\ref{fig:autoGRF} displays the peak and filament auto-correlation and Figure~\ref{fig:pf-abundancesGRF} the peak-filament cross-correlation function for various abundances. To get these results, we employed a power-law power spectrum with spectral index $n_s=-1.5$ which is close to a $\Lambda$CDM power spectrum at the scales probed in this paper. In principle, similar calculations -- although more computing-time consuming -- could be performed in this case but would not change significantly the result.
 The   ${\cal P\!F}$, ${\cal F\!F}$ and ${\cal P\!P}$ 2pCFs are found to display a maximum  at a few smoothing lengths  (with decreasing amplitude in this order)  before decreasing asymptotically towards $0$ at large separations.  The $\cal P\!F$ diverges as $r\to0$, as expected since one can find infinitely close peak-filament pair without having unlikely constraints to fulfil such as a critical point of another kind in between. This is not the case for ${\cal P\!P}$ (resp. ${\cal F\!F}$) which have to go to $-1$ at zero separation

 as one needs to go through a filament-saddle (resp. a peak or a wall-type saddle) between the two components of the pair. 
 On the other hand, the ${\cal P\!W}$,  ${\cal P\!V}$ and  ${\cal F\!W}$ 2pCFs  are 
almost identical: they are always negative, with an extended exclusion zone at small separations followed by a monotonic growth towards $\xi=0$ at infinity. 
The respective extent of their exclusion zone 
are ordered in the same way as in the simulation.   All correlations show qualitatively consistent features with the measured 2pCFs at redshift zero (Figure~\ref{fig:auto_h},~\ref{fig:cross1_h} and \ref{fig:cross2_h}).

 For auto 2pCF as a function of the abundance (Figure~\ref{fig:autoGRF}), the  shape and amplitude match  qualitatively the non-linearly evolved correlations presented in the main text (Figure~\ref{fig:auto_h}, top and middle left panels). As expected, the rarer the critical points, the more biased the tracers and the larger the amplitude of the 2pCF. The same conclusion holds for the ${\cal P\!F}$ 2pCF on Figure~\ref{fig:pf-abundancesGRF}, which should be compared to the top left panel of Figure~\ref{fig:cross2_h}.

In all cases, we note that the predicted correlation function is always larger for rarer critical points, a feature consistent with the findings on evolved simulations reported in the main text, which in itself is remarkable, given the level of non-linearity probed ($\sigma\sim 0.6$).
Also note that for GRFs, rare peaks (above a given threshold) have the exact same behaviour as rare void (below the same threshold with opposite sign) and rare filaments are equivalent to rare walls (due to the field symmetry $\delta\rightarrow-\delta$) which is why we only display a subset of all correlation functions, the rest being redundant (for instance $\cal V\!V$ is identical to $\cal P\! P$).

\begin{figure*}
\add\centering	\includegraphics[width=0.85\columnwidth]{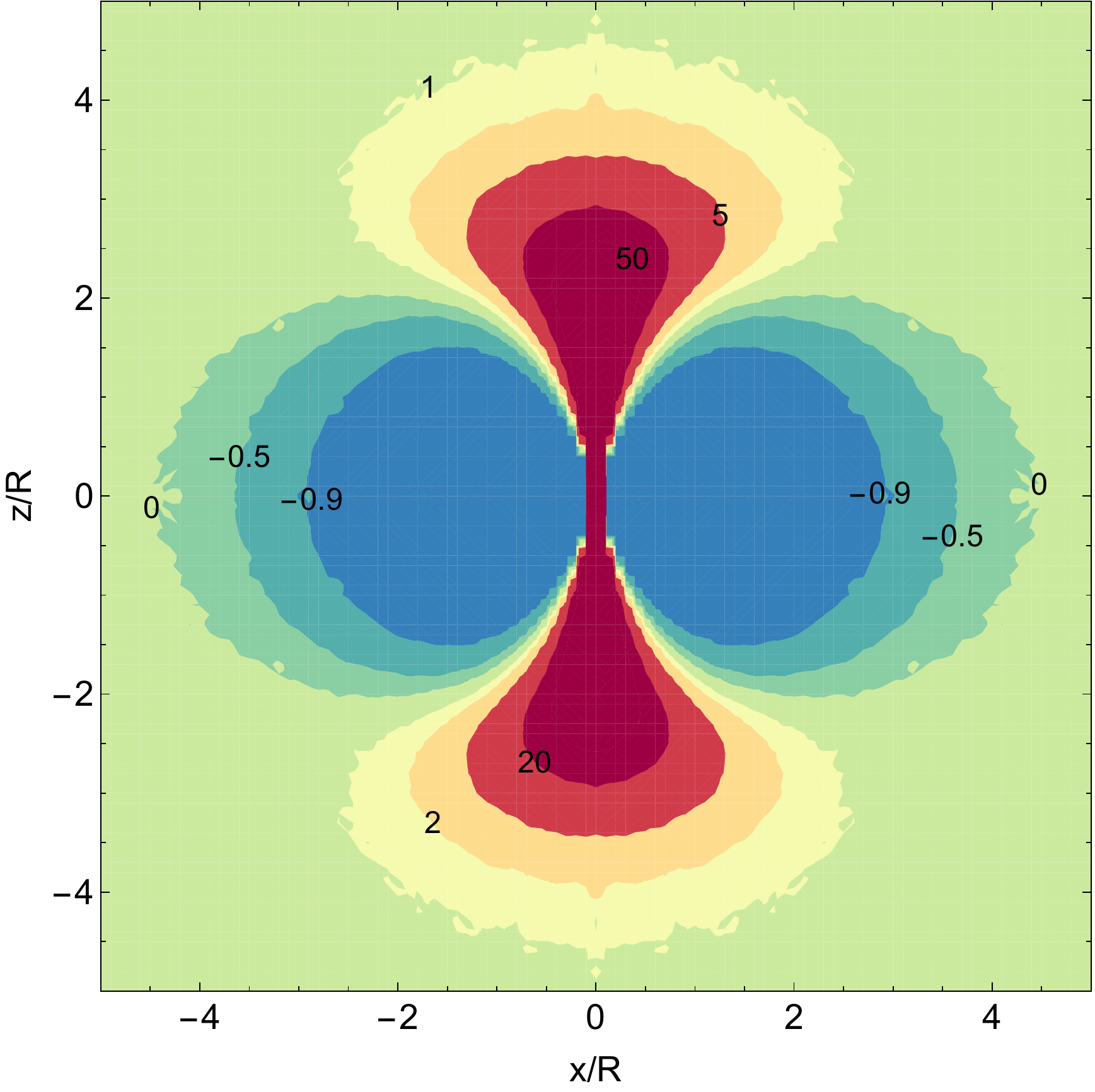}
\includegraphics[
width=1.1\columnwidth]{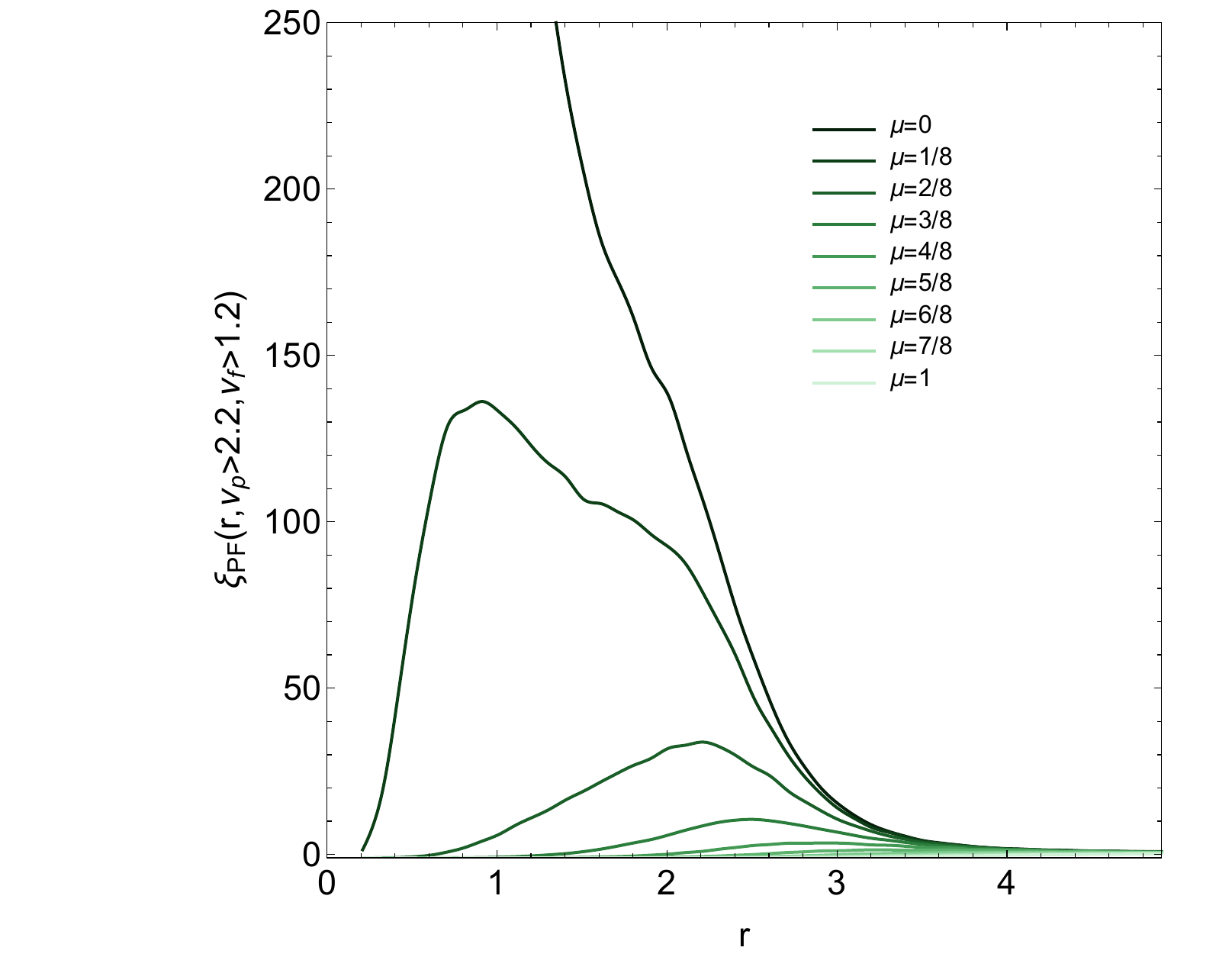}
    \caption{
    {\sl Left panel}: Angle dependence of the peak-filament cross-correlation for GRF in the frame of a saddle for
    the 20\% rarest peaks and filaments. The origin is the location of the saddle point, (Ox) the direction of the void (largest negative curvature) and (Oz) the direction of the filament (positive curvature). This slice corresponds to $y=0$.
    {\sl Right panel}: Slices of the left panel along different directions as labelled.
    Here $\mu$ is the cosine of the angle measured w.r.t. the $x$ axis.
     This figure can be compared  favourably to redshift zero fields in the left panel of Figure~\ref{fig:3dpeakcount} and the top left panel of Figure~\ref{fig:PFevec}.
    }
    \label{fig:crossGRF2D}
\end{figure*}

\begin{figure*}
\add\centering	\includegraphics[trim=0 0 0 0,clip,width=0.9\columnwidth]{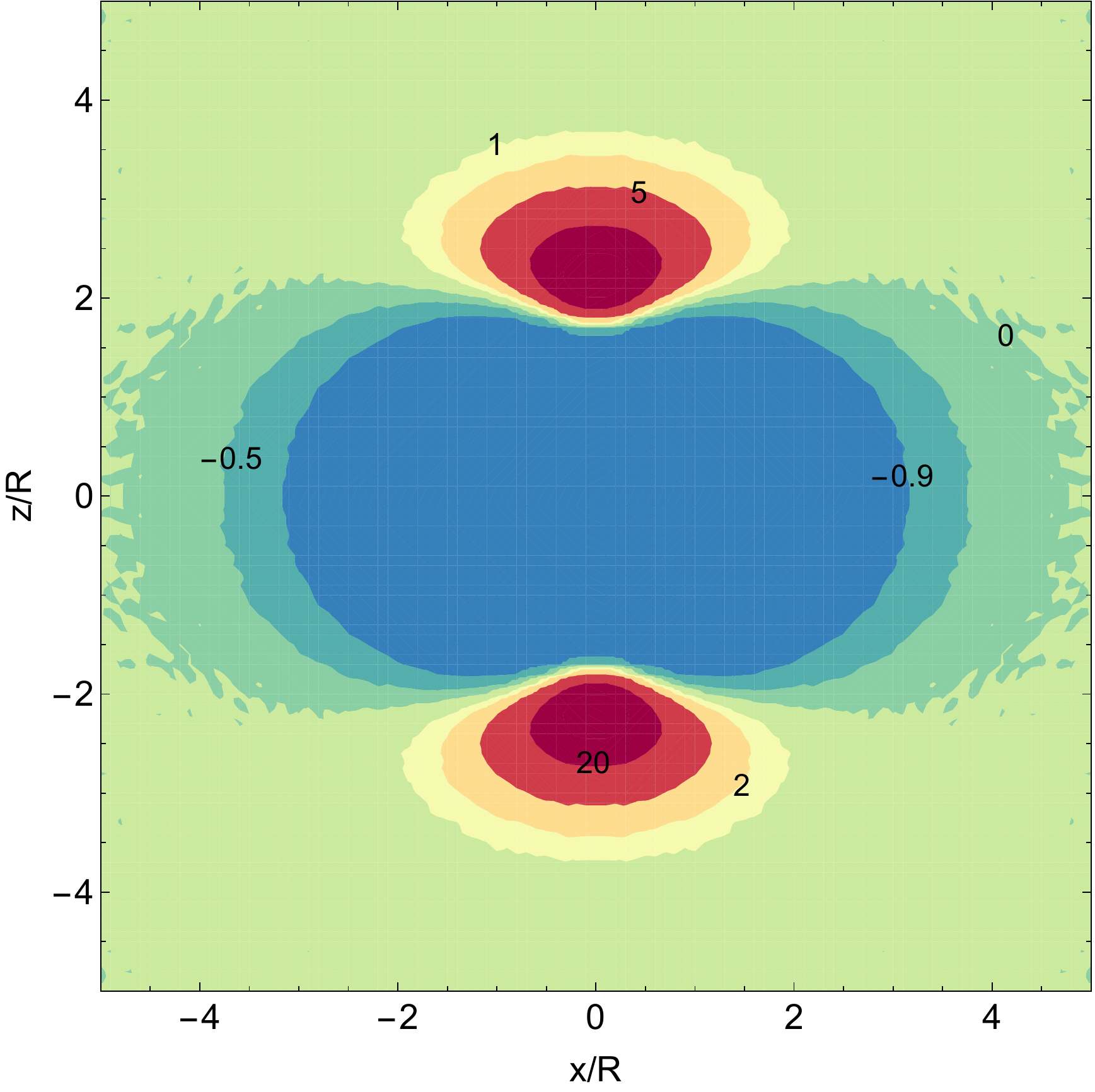}
\includegraphics[trim=0 0 0 0,clip,width=1\columnwidth]{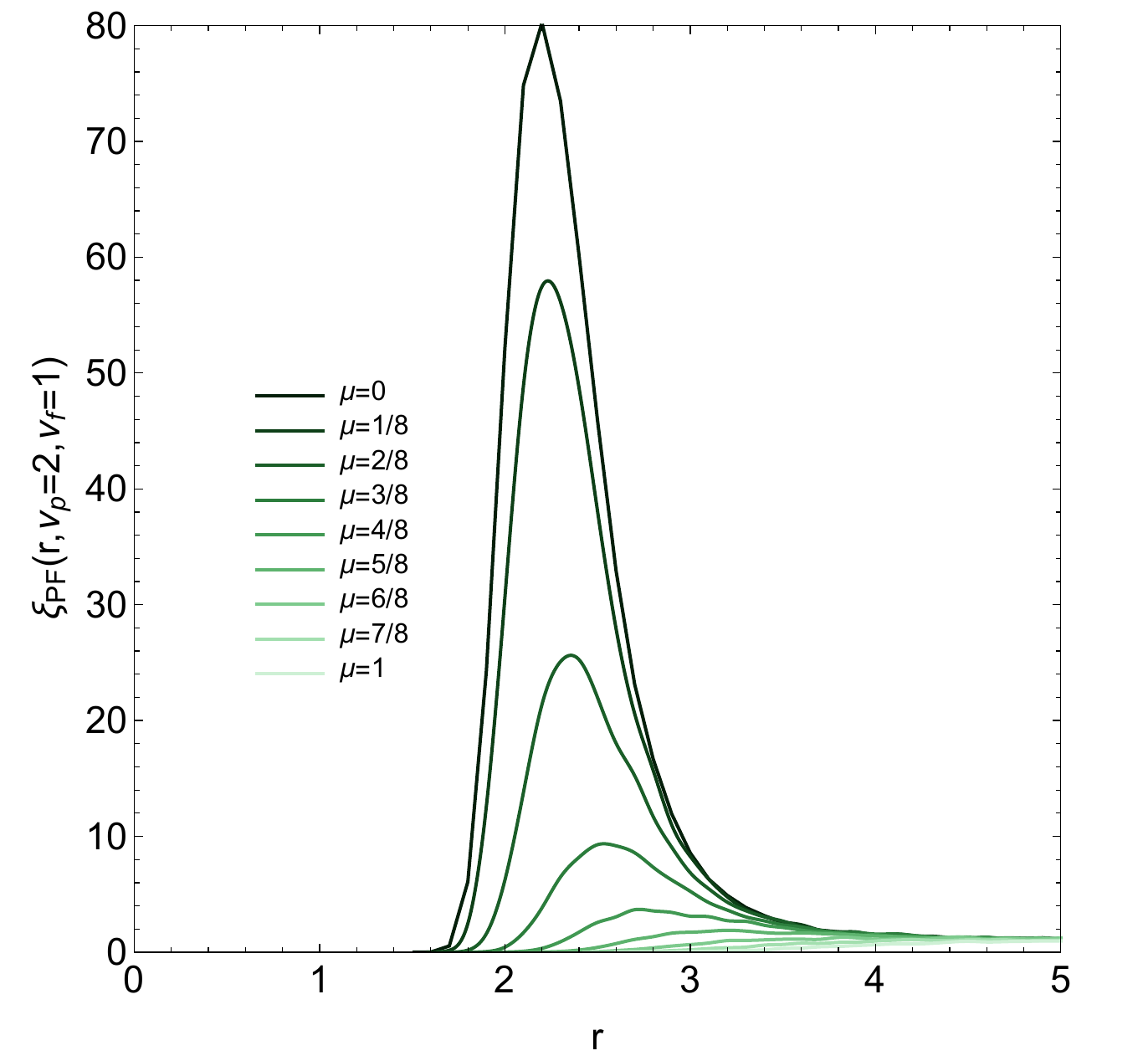}
    \caption{
    {\sl Left and right panel}: Same as Figure~\ref{fig:crossGRF2D}  in the frame of a saddle of rarity $\nu_f=1$ when peaks are set to have a rarity $\nu_p=2$. Since the rarity do not overlap anymore, a global exclusion appears for all orientations,
    though the distribution of peaks remains strongly anisotropic.
     }
    \label{fig:crossGRF2D-nuth}
\end{figure*}

\subsection{Oriented clustering}

Figure~\ref{fig:crossGRF2D} then shows predictions for the ${\cal P\!F}$ cross-correlation in the frame of the 
saddle using GRF. For that purpose, we adopt the same algorithmic strategy
 as described in Regaldo-Saint Blancard et al., in prep but replacing the orientation of the tidal tensor used there by the orientation of the density Hessian. 
 This is achieved by adding in Equation~(\ref{eq:cond2pt}) the constraint
\begin{multline}
{\cal B}(\boldsymbol{X})=2\pi^2(x_{33}-x_{22})(x_{22}-x_{11})(x_{33}-x_{11})\delta_{\rm D}(x_{12})\\
\times\delta_{\rm D}(x_{13})\delta_{\rm D}(x_{23})
\Theta_{\rm H}(x_{33}-x_{22})\Theta_{\rm H}(x_{22}-x_{11}),
\end{multline}
which allows us to impose the off-diagonal coefficients of the density hessian to be zero, the curvatures to be ordered and to add the Jacobian of the transformation from arbitrary frame to the frame of the density Hessian by means of the usual Vandermonde determinant $(x_{33}-x_{22})(x_{22}-x_{11})(x_{33}-x_{11})$ with an additional $2\pi^2$ due to the integration over the Euler angles.

 The result for the ${\cal P\!F}$ cross-correlation for the 20\% highest points in the frame of the 
saddle is displayed in Figure~\ref{fig:crossGRF2D} for again a spectral index $n_s=-1.5$ and the separation is shown in units of the (arbitrary) smoothing length.
 The resolution is 10\% of the smoothing length, hence the  artefact close to the origin along the filament direction (Oz).
Interestingly, the qualitative shape and amplitude of these functions is very similar to what has been measured in gravitationally evolved simulations of the large-scale structure with a correlation bridge between the two peaks on both side (up and down) of the saddle point (at the origin of the frame), and exclusion zones in the perpendicular directions.
In contrast, we also plot the ${\cal P\!F}$ cross-correlation for critical points of fixed heights (1 for filaments and 2 for peaks in this example) in Figure~\ref{fig:crossGRF2D-nuth}. The exclusion zones in the transverse directions remain with the two peaks along the filament axis, but the correlation bridge between the peaks with the divergence at zero separation is replaced by an exclusion zone due to the different height constraint, which does not allow the peak and saddle (of different height) to stand too close to each other.

%%%%%%%%%%%%%%%%%%%%%%%%%%%%%%%%%%%
\subsection{BAO scale predictions}
\label{sec:BAOtheory}

On BAO scales, expansion bias theory \citep[\eg][and reference therein]{Desjacques2018} allows us to anticipate the sign of the amplification on these scales. 
At leading order the peak two-point function is described by the first scale dependent peak bias \citep{desjacques08,2010PhRvD..82j3529D}
\begin{align}
\xi^{(\rm LO)}_{\cal PP}\approx  &\ \bar b_{10}(\nu_1)\bar b_{10}(\nu_2)\xi_{0,0}+
\bar b_{01}(\nu_1)\bar b_{01}(\nu_2)\xi_{4,0}
\notag\\
&+[\bar b_{10}(\nu_1)\bar b_{01}(\nu_2)+\bar b_{01}(\nu_1)\bar b_{10}(\nu_2)]\xi_{2,0}\,, \label{eq:xiexpand}
\end{align}
where the
$\xi_{i,j}$ are defined as
\begin{equation}
\xi_{i,l}=\int \frac{\dd^3 k}{(2\pi)^3} P_s(k) k^i j_l(k r)\; ,
\end{equation}
with $j_l$ the spherical Bessel functions of order $l$,
and the bias factors $\bar b_{ij}$ are given by
\begin{equation}
\bar b_{10}(\nu,x) =\frac{1}{\sigma_0}\frac{\nu-\gamma \bar u}{1-\gamma^2}\,,\quad
\bar b_{01}(\nu,x) =\frac{1}{\sigma_2}\frac{\bar u-\gamma \nu}{1-\gamma^2}\,,
\end{equation}
with the mean peak curvature $\bar u =-\left\langle x_{11}+x_{22}+x_{33} |x=\nu\right\rangle$.
In equation~\eqref{eq:xiexpand}, the $\xi_{0,0}$ term correspond to the linear local bias which dominates for very high peak thresholds. This contribution which only comes from the peak height constraint \citep{BBKS,desjacques08} is typically negative for underdense regions and positive for overdense regions scaling like $\sigma_0 b_{10}(\nu)\approx \nu -3/\nu$ at high $\nu$\footnote{ 
Note that an extension of the predicted linear bias to the quasi-linear regime was performed in \citet{Uhlemann:2016un}.}.
Since we can anticipate this sign for a given critical type and choice of abundance, and given that correlations involve 
the product of  bias factors in equation~\eqref{eq:xiexpand}, all observed bumps and dips
in Figures~\ref{fig:auto_h}-\ref{fig:cross2_z} are qualitatively  consistent with expansion bias theory. The contribution from 
the higher order derivatives, $\xi_{2,0}$ and $\xi_{4,0}$ typically tend to amplify and sharpen the BAO signature which tends to be damped by the smoothing operation, and it also reduces the inward shift of the BAO peak due to this smoothing (the BAO wiggles being asymmetric). However, the combined effect of those three terms occurs
in some non trivial manner depending on height, filter and scale in particular. Note that higher order bias expansion have been performed  \citep{2013PhRvD..87d3505D,matsubara&codis19} but are negligible to describe BAOs in Lagrangian space. 
Let us emphasize that the subsequent gravitational evolution will tend to broaden the initial sharp BAOs due to velocity drift. This effect can be investigated by modelling the Lagrangian displacement of initial peaks with perturbation theory \citep{2010PhRvD..82j3529D} or more appropriately in our context using a Gram-Charlier expansion at the level of the joint statistics as was done on one-point statistics in e.g \cite{Gay2012}. This is left for future works.

\subsection{Cosmic crystal for GRF} \label{sec:crystal}
%%%%%%%%%%%%%%%%%%%%%%%%%%%%%%%%%%%%%%%%%%%%%%%%%%

\begin{figure}
	\includegraphics[trim=0 0 0 0,clip,width=1.\columnwidth]{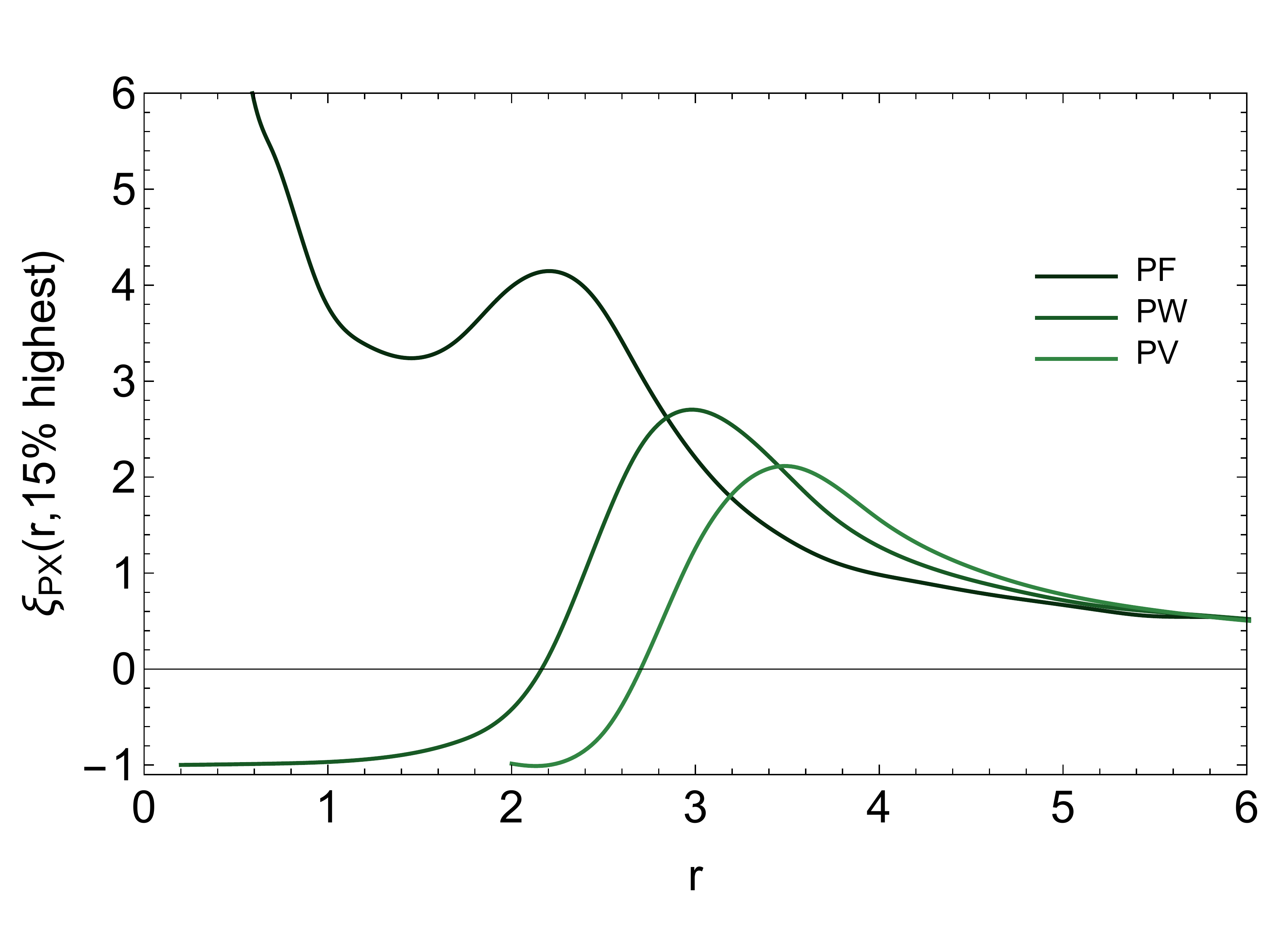}
 \centering	\includegraphics[trim=0 0 0 0,clip,width=0.85\columnwidth]{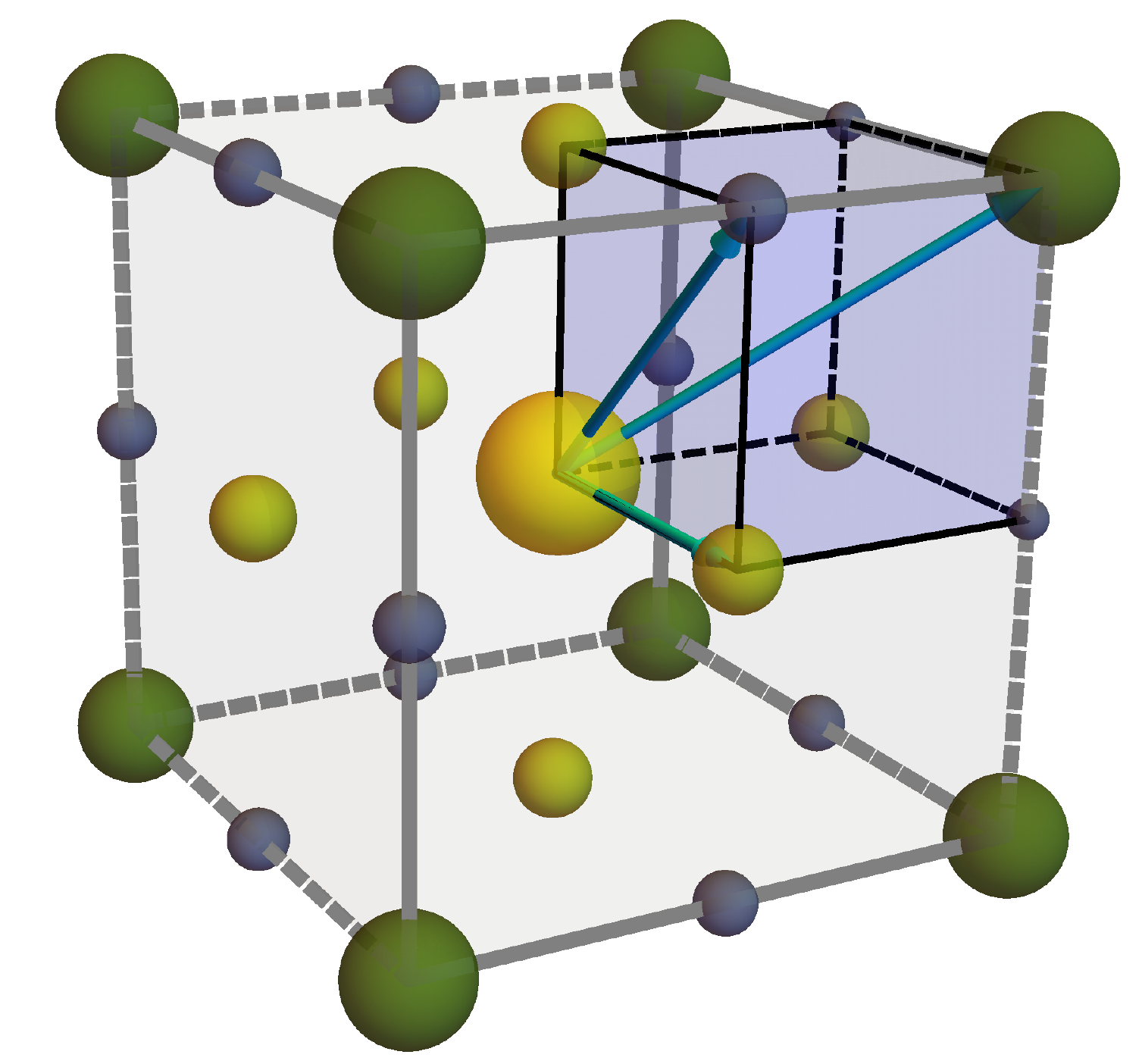}
    \caption{{\sl Top panel:} cross-correlation  between peaks and 
 the 15\% highest critical points for a  GRF with $n_s=-1.5$. The ratio of the radii at which these correlation reach their maxima are given by $r_{\cal P\!F}/r_{\cal P\!W}=0.7\approx 1/\sqrt{2}$,   and  $r_{\cal P\!F}/r_{\cal P\!V}=0.6\approx 1/\sqrt{3}$ , which in turn suggest  that the cosmic web has  (on average) a cosmic cubic centred crystal structure,  as shown on the {\sl bottom panel}. 
 Here, the peak is represented as a big yellow sphere, the filament saddle as  green spheres, the  wall saddles as small blue spheres, and the voids as big  light green spheres. The blue arrows lengths are resp. $r_{\cal P\!F} $, $r_{\cal P\!W}$ and  $r_{\cal P\!V}$. 
 }
    \label{fig:crystal}
\end{figure}

The peak-filament, peak-wall and peak-void correlation functions all present a maximum that we can use as a reference for the typical position of the first layer of respectively filaments, walls and voids around a given peak. For GRF, we compute those correlation functions for the $15\%$ highest critical points and display the result on Figure~\ref{fig:crystal}, top panel. 
The corresponding ratios are in agreement with those presented at redshift zero in the main text (Figure~\ref{fig:crystal_measure}).
Figure~\ref{fig:crystal}, bottom panel, shows a qualitative rendering of the corresponding crystal, consistent with these ratios.

\section{Cross-correlation divergence} \label{sec:cross-rare}
%%%%%%%%%%%%%%%%%%%%%%%%%%%%%%%%%%%%%%%%%%%%%%%%%%

Let us finally address the origin of the  difference  in shape between  correlations involving saddle points.
In the main text, Figure~\ref{fig:cross2_h} displayed some divergence at this origin,  whereas Figure~\ref{fig:cross1_h} did not.
It turns out that the lack of divergence at the origin of  $\cal F\!W$ correlation only reflects  the  rarity difference  between the chosen filaments and walls.
\begin{figure*}
	\includegraphics[trim=0 15bp 0 15bp,clip,width=1.\columnwidth]{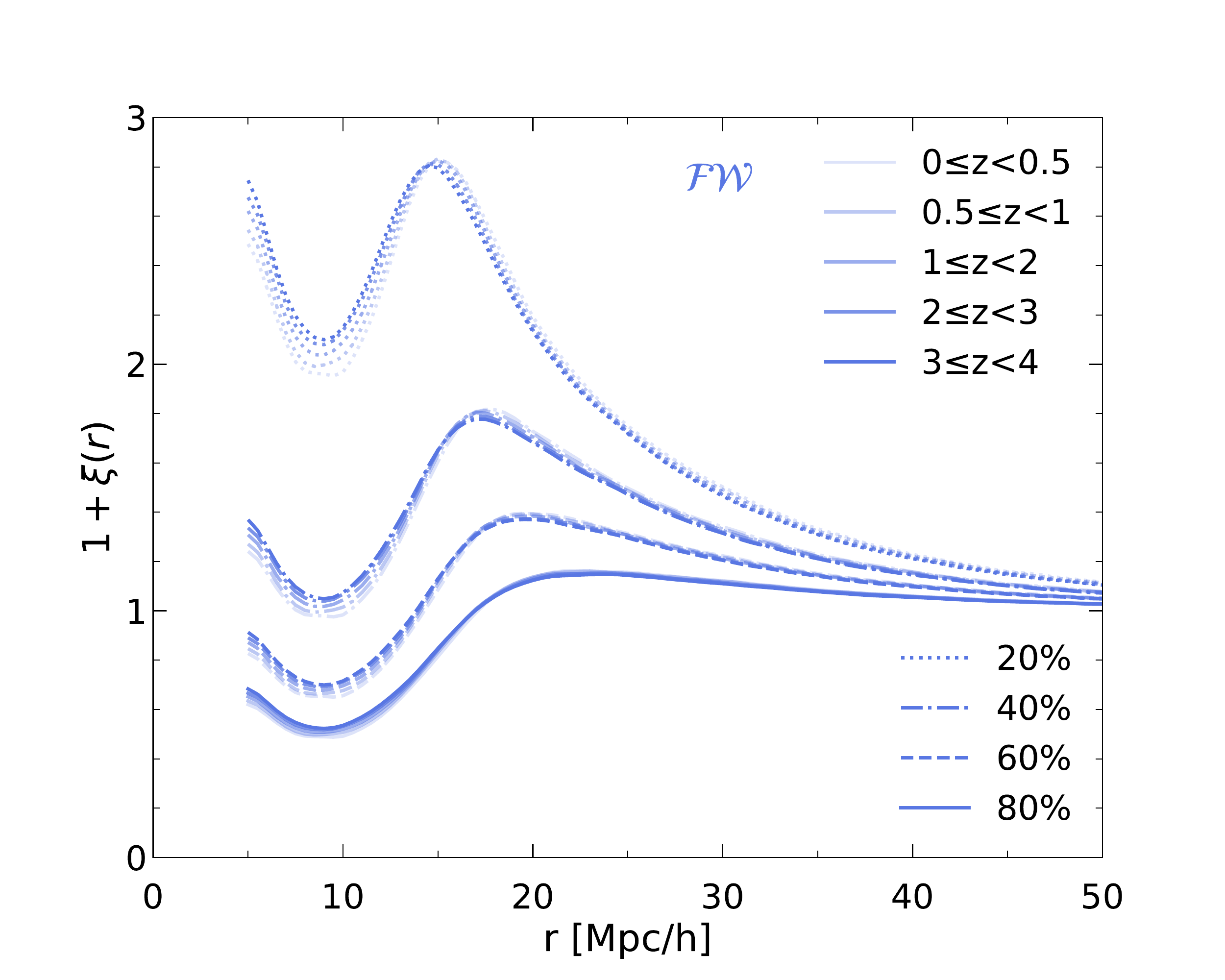}
		\includegraphics[width=\columnwidth]{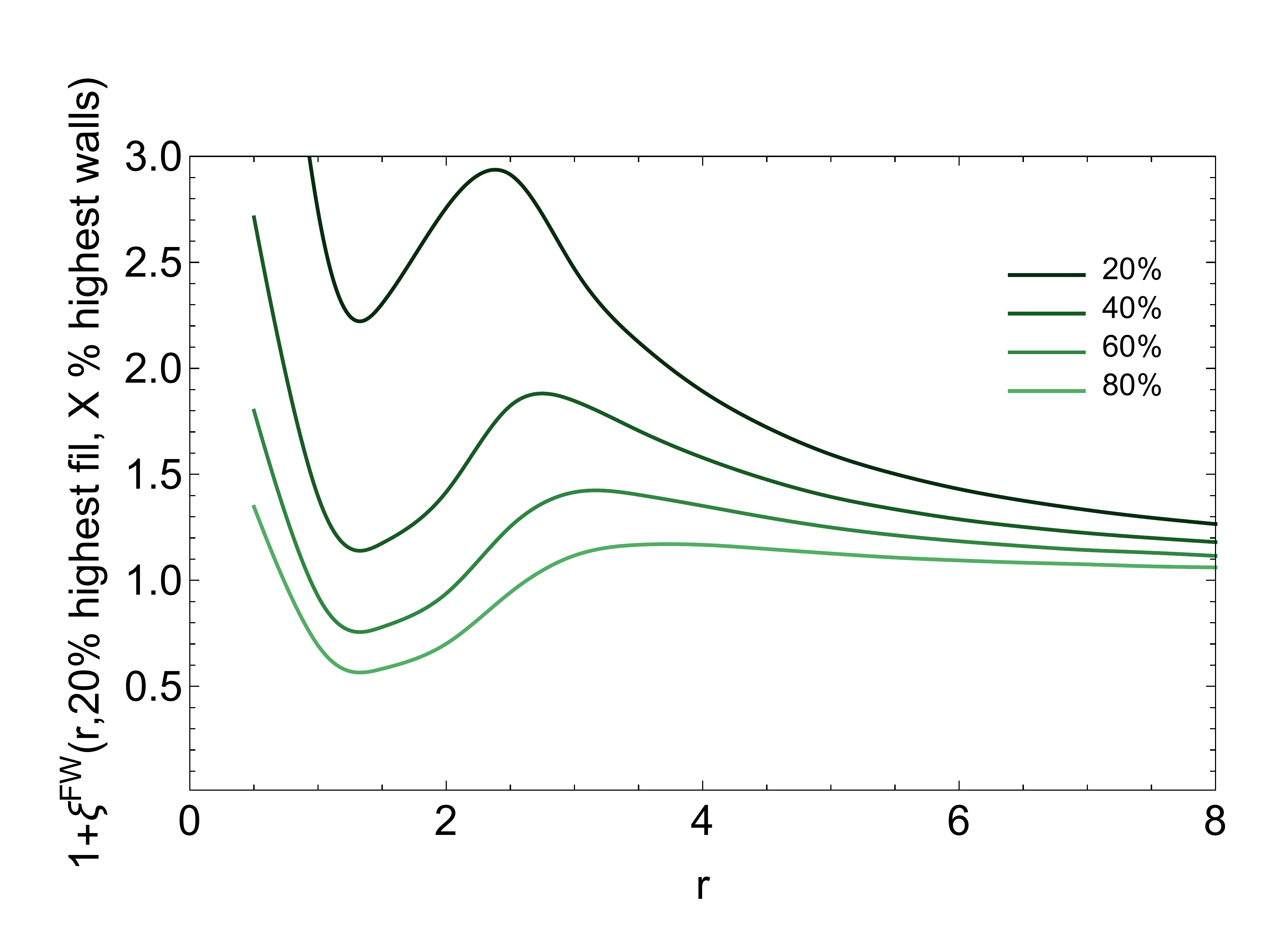}
    \caption{{\sl Left panel:} The behaviour of the $\cal F\!W$ correlation at small separation as a function of the abundance of wall-points above a certain rarity threshold. For walls,
it varies here between 20\% and 80\%, whereas  filaments are fixed at 20\%. From the perspective of 
    critical mergers \citep{cadiou2020}, it is expected that filament and wall saddles  merge during  cosmic evolution, which was {not seen}
    on Figures~\ref{fig:cross1_h} and \ref{fig:cross1_z}. This was due to the chosen rarity difference between filaments and walls. As we increase here the rarity (including higher walls) the exclusion zone disappears and we recover the qualitative shape of the $\cal P\!F$ and $\cal W\!V$ correlations (Figures~\ref{fig:cross2_h} and \ref{fig:cross2_z}), including the divergence at small separation. Note also the  stability of these correlation with redshift.
    {\sl Right panel:} same as left panel for GRF. The separation is in units of the Gaussian smoothing length. The  qualitative similarity between the left and right panels is 
    striking. }
    \label{fig:cross_FW+}
\end{figure*}
%%%%%
Indeed Figure~\ref{fig:cross_FW+} now presents the $\cal F\!W$ correlation as a function of separation when the abundance above (in contrast to below as was used in the main text) a threshold rarity for wall-points is varied from 20\% to 80\%, whereas that for filament-points is fixed to 20\%. As expected, when the abundances are inconsistent, an exclusion zone occurs, whereas when they allow for some overlap, one recovers a divergence, as expected from critical event theory \citep{cadiou2020}: filament and wall saddle points do get close to each other when a wall disappears.     
Note interestingly that these cross-correlations  are quite insensitive to redshift evolution.

%%%%%%%%%%%%%%%%%%%%%%%%%%%%%%%%%%%%%%%%%%%%%%%%%%
\noindent

% Don't change these lines
\bsp	% typesetting comment
\label{lastpage}
\end{document}